\documentclass[aps,10pt,notitlepage,nofootinbib,superscriptaddress]{revtex4-2}

	\usepackage{graphicx,color}
	\usepackage{amsmath}
	\usepackage{amssymb}
	\usepackage{hyperref}
	\usepackage[utf8]{inputenc}
	\usepackage[english]{babel}
	\usepackage{epsfig}
	\usepackage{subfigure}
	\usepackage{wasysym}
	\usepackage{color,xcolor}
	\usepackage{amsmath}
	\usepackage{bm}
	\usepackage{epsfig}
	\usepackage{amsfonts}
	\usepackage{dcolumn}
	\usepackage{float}
	\usepackage{booktabs}
	\usepackage{relsize}
	
\begin{document}

\title{Black Bounce Solutions from a Self-Interacting 3-Form Field in General Relativity}

\author{Francisco S. N. Lobo} \email{fslobo@ciencias.ulisboa.pt}
\affiliation{Institute of Astrophysics and Space Sciences, Faculty of Sciences, University of Lisbon, Building C8, Campo Grande, P-1749-016 Lisbon, Portugal}
\affiliation{Department of Physics, Faculty of Sciences, University of Lisbon, Building C8, Campo Grande, P-1749-016 Lisbon, Portugal}

\author{Manuel E. Rodrigues} \email{esialg@gmail.com}
\affiliation{Faculty of Physics, Graduate Program in Physics, Federal University of Pará, 66075-110, Belém, Pará, Brazil}
\affiliation{Faculty of Exact Sciences and Technology, Federal University of Pará, Abaetetuba University Campus, 68440-000, Abaetetuba, Pará, Brazil}

\begin{abstract}

We construct a new class of black-bounce solutions sourced by a self-interacting 3-form field minimally coupled to general relativity and a scalar field. The 3-form field, which naturally arises in string theory, supergravity, and cosmological models, provides the anisotropic effective stresses required to sustain regular geometries that interpolate smoothly between black holes and traversable wormholes. By exploiting the Hodge duality between a 3-form and a 1-form in four dimensions, we reduce the field equations and obtain exact solutions through the direct integration of the coupled equations of motion. In particular, the solutions are derived from algebraic combinations and manipulations of the Einstein, scalar, and 3-form field equations, starting from a complete action principle, without employing the usual reconstruction procedure in which the metric ansatz is imposed a priori and the matter sector is reconstructed afterwards. This approach reveals two distinct classes of solutions. The first one yields a globally phantom scalar field and a metric function with a characteristic arctangent dependence, reducing to the Schwarzschild--(anti) de Sitter spacetime in the limit of vanishing 3-form coupling. The second class produces a constant 3-form Lagrangian and, remarkably, a partially canonical scalar field, namely phantom only near the bounce and canonical outside the event horizon, a feature previously attainable mainly in modified theories of gravity, but which emerges here within pure general relativity. Both families are globally regular, as confirmed by the finiteness of the Kretschmann scalar, and exhibit an asymmetric horizon structure inherited from the 3-form energy-density distribution. To assess their phenomenological viability, we constrain the bounce parameter using the observations of the shadow of Sagittarius A$^\ast$ by the Event Horizon Telescope collaboration. For the first solution, compatibility with the $2\sigma$ bounds requires $0 \leq h_0/M \leq 1.49,$ while for the second solution the allowed range is $629.56 \leq h_0/M \leq 1117.29.$ These results demonstrate that the 3-form black-bounce framework is both mathematically consistent and observationally viable, offering a compelling alternative to standard black-hole paradigms with testable predictions for future high-precision astrophysical observations.

\end{abstract}

\date{\today}
\maketitle
\def\HMS{{\scriptscriptstyle{\rm HMS}}}

\section{Introduction}

Antisymmetric tensor fields occupy a central position in modern theoretical physics, appearing in gravitational theories, quantum field theory, cosmology, supergravity, and string theory. Among these objects, $3$-form fields are particularly interesting because they can act as non-trivial gravitational sources and generate a wide variety of geometrical, topological, and cosmological phenomena. Their versatility has motivated extensive investigations ranging from fundamental high-energy theories to phenomenological models of the Universe and compact astrophysical systems.

The modern development of antisymmetric tensor fields can be traced back to the pioneering work of Kalb and Ramond~\cite{Kalb:1974yc}, who introduced the antisymmetric tensor field $B_{ab}$ in the context of string theory. Its field strength, $H_{abc} = \partial_a B_{bc} + \partial_b B_{ca} + \partial_c B_{ab}$, defines a $3$-form, commonly known as the Kalb--Ramond field, which later became a fundamental component of the Neveu--Schwarz--Neveu--Schwarz (NS--NS) sector of superstring theories~\cite{Green:2012oqa,Polchinski:1998rq}. Antisymmetric tensor fields also appear naturally in topological gravity and supergravity. Their role in eight-dimensional topological gravity was explored in Ref.~\cite{Baulieu:2003my}, while eleven-dimensional supergravity incorporates a fundamental $3$-form potential $A_{abc}$ with associated field strength $F_{abcd}=4\partial_{[a}A_{bcd]}$, which plays a crucial role in M-theory compactifications and M2-brane dynamics~\cite{Cremmer:1978km}. More recently, Kalb--Ramond sectors have been shown to contribute to axion quality protection in string-inspired effective theories, highlighting the interplay between ultraviolet physics and low-energy phenomenology~\cite{Burgess:2023ifd}.

The cosmological applications of $3$-form fields have attracted considerable attention. Such fields provide viable frameworks for inflation, dark energy, and bouncing cosmologies~\cite{Koivisto:2009sd,Koivisto:2009fb,Mulryne:2012ax}. Canonical minimally coupled theories admit isotropic solutions exhibiting scaling behaviour, transient acceleration, and phantom-crossing phenomena, with the cosmological dynamics being conveniently described through an effective potential formalism~\cite{Koivisto:2009ew}. Inflationary models driven by three-form fields have been studied extensively, including analyses of cosmological perturbations, non-Gaussianities, reheating, and the stability conditions required to avoid ghosts and Laplacian instabilities for various classes of potentials~\cite{DeFelice:2012jt,Urban:2012ib,DeFelice:2012wy}. Additional investigations have considered anisotropic inflation generated by couplings between three-form and Abelian gauge fields, which can produce substantial anisotropy and thereby provide observational constraints on such interactions~\cite{Urban:2013aka}. Multifield scenarios involving two $3$-form fields have been explored, revealing complex interactions mediated by the Hubble parameter, the generation of isocurvature perturbations, and a varying sound speed~\cite{Kumar:2014oka}. The primordial non-Gaussianity produced in multiple three-form inflation has been computed using the $\delta N$ formalism, yielding results compatible with present observational constraints, including $\mathcal{O}(1)$ amplitudes in equilateral and orthogonal configurations~\cite{SravanKumar:2016biw}.

Interacting three-form dark-energy models have likewise been proposed as viable descriptions of the late-time Universe, with the aim of distinguishing phenomenological interactions through statefinder hierarchy parameters and growth factor diagnostics while avoiding abrupt late-time events such as the Little Sibling of the Big Rip~\cite{Morais:2016bev}. Non-minimal couplings between $3$-form fields and gravity have also generated substantial interest. Such couplings can give rise to effective cuscuton descriptions, where a scalar field with no dynamical degrees of freedom emerges naturally from the three-form sector~\cite{DeFelice:2025khe}. Anisotropic cosmological solutions sourced by Kalb--Ramond backgrounds have been constructed in Bianchi type I geometries~\cite{Maluf:2021eyu}, while bouncing cosmologies in generalized teleparallel gravity have been studied in the presence of antisymmetric tensor sectors~\cite{Karthikeyan:2021vpk}. Within braneworld contexts, Kalb--Ramond fields have been shown to contribute to the regularisation of cosmological singularities~\cite{DeRisi:2008qw}. Additional extensions have explored parity-violating gravitational sectors, dynamical dark-energy behaviour, and topological corrections to gravity arising from non-minimal couplings between Kalb--Ramond fields and spacetime geometry~\cite{Manton:2024hyc,Hell:2026mle,Capanelli:2023uwv}.

From a geometrical perspective, $3$-form fields possess several remarkable properties. In four spacetime dimensions, the Hodge dual of a $3$-form is a vector field, whereas the dual of its field strength is a pseudoscalar. Consequently, these theories establish direct connections with axion physics, effective pseudoscalar models, and torsion-based gravitational frameworks. Such features have motivated the study of compact gravitational systems supported by antisymmetric tensor sectors. Particularly interesting applications arise in black-hole, wormhole, and compact-star physics. Wormhole geometries sustained by three-form curvature terms have been shown to satisfy the null and weak energy conditions for the ordinary matter threading the spacetime, with all exoticity confined to the three-form sector~\cite{Barros:2018lca}. Static three-form stars have also been investigated and found to possess larger masses than comparable general-relativistic configurations, suggesting possible applications to compact objects such as the secondary component of GW190814~\cite{Barros:2021jbt}.

Traversable wormholes supported by massive self-interacting three-form fields have furthermore been constructed within pure Einstein gravity, yielding regular asymptotically flat geometries with a single throat whose shadows may mimic those of black holes when the self-interaction is sufficiently large~\cite{Bouhmadi-Lopez:2021zwt}. Three-form Lagrangians capable of reproducing both matter-dominated and dark-energy epochs have been developed through dynamical-systems methods, identifying viable scenarios that replicate a de Sitter universe in the present epoch~\cite{daFonseca:2024boz}. Related studies involving massive vector fields minimally coupled to Einstein gravity have revealed additional mechanisms for sustaining wormhole geometries within effective field theories, where the existence of at least one ghost degree of freedom provides the necessary conditions to support these exotic configurations~\cite{Ganiyeva:2024boz}. On the observational side, three-form dark-energy models with Gaussian potentials have recently been confronted with Planck PR4 cosmic microwave background data, DESI DR1 baryon acoustic oscillation measurements, Pantheon+ Type Ia supernovae data, and large-scale structure observations~\cite{Bouhmadi-Lopez:2025lzm}. These analyses indicate that three-form dark energy can provide a viable phantom-like component while mildly alleviating current cosmological tensions. Collectively, these developments establish $3$-form fields as versatile ingredients connecting string theory, supergravity, cosmology, modified gravity, and compact-object physics.

In the present work, we investigate a new class of black-bounce (BB) geometries supported by a self-interacting $3$-form sector minimally coupled to general relativity and a scalar field. The black-bounce paradigm introduced by Simpson and Visser~\cite{Simpson:2018tsi} replaces the central singularity of a black hole by a regular throat of finite radius, thereby interpolating continuously between regular black holes and traversable wormholes. We demonstrate that the intrinsic anisotropic stresses generated by a self-interacting $3$-form field naturally support such geometries, providing a physically motivated alternative to the non-linear electrodynamics sources frequently employed in the literature. The theory is governed by a self-interaction Lagrangian $\mathcal{L}(H_{2})$, where $H_{2}=H_{abc}H^{abc}$. Using the Hodge-dual representation of the three-form field, we reduce the field equations to a tractable system of ordinary differential equations.

A central methodological novelty of this work is that the black-bounce solutions are obtained through the direct integration of the equations of motion, rather than by prescribing the geometry and subsequently reconstructing the matter sector to support it. In the standard BB framework, the metric functions are typically postulated \emph{a priori}, most notably the Simpson--Visser profile $\Sigma(r)=\sqrt{r^{2}+h_{0}^{2}}$, and the Einstein equations are then solved algebraically for the required matter content. While computationally convenient, this reconstruction approach leaves open the question of whether the resulting matter fields arise naturally from a well-defined fundamental action. By contrast, the procedure developed here starts from a complete action principle comprising the $3$-form field, the scalar field, and their self-interactions, and integrates the coupled field equations directly. The solution emerges from the dynamics of the theory itself, without any \emph{ad hoc} geometric ansatz. To the best of our knowledge, this constitutes the first example of a BB solution in general relativity that is derived in this fully self-consistent manner, establishing a new level of rigor in the construction of regular bouncing geometries.

This procedure reveals two distinct branches of solutions associated with different realizations of the consistency condition required for $\mathcal{L}$ to be a genuine function of $H_{2}$.
The first branch yields a globally phantom scalar field with kinetic coupling $\epsilon=-1$, a linear three-form Lagrangian, and the standard Simpson--Visser areal radius $\Sigma(r)=\sqrt{r^{2}+h_{0}^{2}}$, where the parameter $h_{0}$ determines the bounce scale. The second branch leads to a constant three-form Lagrangian and a partially canonical scalar field that is phantom only near the bounce and becomes canonical outside the event horizon. This configuration requires a deformed areal radius controlled by an additional parameter $k_{1}$ and reproduces a behaviour previously encountered only within modified theories of gravity. Both classes of solutions are free of curvature singularities, possess finite Kretschmann scalars throughout the spacetime, and admit asymptotically de Sitter, anti-de Sitter, or asymptotically flat limits depending on the value of the cosmological constant $\Lambda$. Their horizon structure is generally asymmetric owing to the underlying three-form energy distribution. To assess their observational viability, we confront the solutions with Event Horizon Telescope observations of the shadow of Sgr A$^{*}$. We find that the first family is compatible with current observational bounds for $0 \le h_{0}/M \le 1.49$, whereas the second family requires $629.56 \le h_{0}/M \le 1117.29$. These results demonstrate that three-form-supported black-bounce geometries are both mathematically consistent and observationally viable, while yielding potentially distinguishable signatures for future high-precision observations.

The paper is organized as follows. In Sec.~II, we introduce the action, derive the field equations, and formulate the consistency condition governing the reconstruction procedure. Sections~III and IV are devoted to the two solution branches and their corresponding black-bounce geometries. In Sec.~V, we derive observational constraints from the shadow of Sgr A$^{*}$ using Event Horizon Telescope measurements. Finally, Sec.~VI summarizes our conclusions and discusses possible directions for future research.

Throughout this work, we adopt geometrized units in which $G=c=1$ and $\kappa^{2}=8\pi$.

\section{Field equations}

In this section, we introduce the action of the theory and derive the equations of motion. The central idea of this work is to employ a self-interacting 3-form field as the matter source in general relativity, with the aim of generating new black-bounce (BB) solutions. We begin by defining the relevant differential forms and their dual representations in four-dimensional spacetime, then construct the action functional, perform the variational procedure, and finally specialise the resulting field equations to static, spherically symmetric geometries.

\subsection{The 3-form field and its dual representation}

Let us define a 3-form field $H$ through its components,
\begin{equation}
	H = \frac{1}{3!}\,H_{abc}\;dx^{a}\wedge dx^{b}\wedge dx^{c}.
\end{equation}
In four dimensions, a 3-form is Hodge-dual to a 1-form. This duality is particularly useful because it allows us to trade the three independent components of $H_{abc}$ for the four components of a vector field $Z^{d}$, subject to one constraint, thereby simplifying the treatment of the 3-form in a spherically symmetric background. We may therefore express the components of $H_{abc}$ in terms of $Z^{d}$ as
\begin{equation}
	H_{abc} = \sqrt{-g}\,\varepsilon_{abcd}\,Z^{d},
	\label{H}
\end{equation}
where $\varepsilon_{abcd}$ is the totally antisymmetric Levi-Civita tensor, with the convention $\varepsilon_{0123}=1$. For a static, spherically symmetric configuration, the vector $Z^{d}$ takes the simple radial form
\begin{equation}
	Z^{d} = \bigl(0,\;\chi(r),\;0,\;0\bigr),
\end{equation}
where $\chi(r)$ is a function of the radial coordinate to be determined by the field equations. The 4-form field strength $F$ associated with $H$ is defined as the exterior derivative $F = dH$, whose components read
\begin{equation}
	F_{abcd} = \nabla_{\!a}H_{bcd} - \nabla_{\!b}H_{cda} + \nabla_{\!c}H_{dab} - \nabla_{\!d}H_{abc}.
	\label{F2}
\end{equation}
In four dimensions, the 4-form $F_{abcd}$ is proportional to the volume form and thus carries a single independent degree of freedom, which will be identified with the kinetic energy of the 3-form field.

For later convenience, we introduce the two independent scalar invariants constructed from the 3-form and its field strength,
\begin{equation}
	H_{2} \equiv H_{abc}H^{abc},\qquad
	F_{2} \equiv F_{abcd}F^{abcd}.
	\label{H2F2}
\end{equation}
The scalar $H_2$ plays a role analogous to the electromagnetic invariant $F_{\mu\nu}F^{\mu\nu}$ in non-linear electrodynamics (NED): it serves as the argument of the self-interaction potential and controls the effective mass and pressure of the 3-form field. The scalar $F_2$, on the other hand, provides the canonical kinetic term, much like the Maxwell term in NED.

\subsection{Action and equations of motion}

The analogy with NED provides a useful guide for constructing the action of the theory. In standard BB solutions sourced by a scalar field and NED, the Lagrangian contains a Maxwell-like kinetic term together with non-linear corrections that depend on the electromagnetic invariant. Here, we associate the 3-form field strength $F_{abcd}$ with the NED field strength $F_{\mu\nu}$, and the scalar $H_2$ with the electromagnetic invariant. This leads us to propose the following action:
\begin{eqnarray}
		S = \int \sqrt{-g}\;d^{4}x\;\Bigg[
		\frac{R}{2\kappa^{2}}
		+ \frac{1}{48}\,F_{2}
		+ \mathcal{L}(H_{2})
		 + V(\varphi) - \epsilon(\varphi)\,\nabla_{\!a}\varphi\,\nabla^{a}\varphi
		\Big],
	\label{action}
\end{eqnarray}
where $R$ is the Ricci scalar and $\kappa^{2}=8\pi G$ (with $G=1$ in geometrised units). The second term, $\frac{1}{48}F_{2}$, is the canonical kinetic term for the 3-form field, with the numerical factor $1/48$ chosen to ensure canonical normalization of the dual vector $Z^{d}$. The function $\mathcal{L}(H_{2})$ is a generic non-linear self-interaction potential for the 3-form, depending only on the invariant $H_2$, whose derivative $\mathcal{L}'(H_{2})$ acts as an effective mass parameter. The scalar field $\varphi$ is included with a potential $V(\varphi)$ and a kinetic coupling function $\epsilon(\varphi)$, whose sign determines whether the scalar field is canonical ($\epsilon>0$) or phantom ($\epsilon<0$). As we shall see, the phantom regime is particularly relevant for sustaining wormhole throats and bouncing geometries.

The functional variation of the action~\eqref{action} with respect to the metric $g^{ab}$ yields the gravitational field equations,
\begin{eqnarray}
		R_{ab} - \frac{1}{2}\,g_{ab}R
		&=& \kappa^{2}\,
		\Bigg[
		\frac{1}{48}\,F_{2}\,g_{ab}
		- \frac{1}{6}\,F_{acde}F_{b}{}^{cde}
			 + g_{ab}\,\mathcal{L}(H_{2})
		- 6\,H_{acd}H_{b}{}^{cd}\,\mathcal{L}'(H_{2}) 
			\nonumber \\
		&& \quad   + g_{ab}\,V(\varphi)
		+ 2\epsilon(\varphi)\Bigl(
		\nabla_{\!a}\varphi\nabla_{\!b}\varphi
		- \frac{1}{2}\,g_{ab}\,\nabla_{\!c}\varphi\nabla^{c}\varphi
		\Bigr)
		\Bigg],
	\label{eqg}
\end{eqnarray}
where the first two terms on the right-hand side constitute the energy-momentum tensor of the 3-form field strength, analogous to the Maxwell stress-energy tensor in NED. The third and fourth terms arise from the self-interaction $\mathcal{L}(H_2)$ and introduce effective anisotropic stresses that are essential for supporting BB geometries, as they allow the radial and tangential pressures to differ in a controlled manner. The fifth and sixth terms are the standard contributions from the scalar field.

The variation with respect to the 3-form components $H^{abc}$ gives the 3-form field equation,
\begin{equation}
	\nabla_{\!d}F_{abc}{}^{d} + 12\,H_{abc}\,\mathcal{L}'(H_{2}) = 0,
	\label{eqH}
\end{equation}
which is a Proca-like equation with an effective mass term proportional to $\mathcal{L}'(H_{2})$. Finally, the variation with respect to the scalar field $\varphi$ yields
\begin{equation}
	2\,\epsilon(\varphi)\,\nabla_{\!a}\nabla^{a}\varphi
	+ V'(\varphi)
	+ \epsilon'(\varphi)\,\nabla_{\!a}\varphi\nabla^{a}\varphi = 0,
	\label{eqphi}
\end{equation}
which reduces to the standard Klein--Gordon equation when $\epsilon(\varphi)$ is constant.

\subsection{Static, spherically symmetric ansatz}

We restrict our analysis to static, spherically symmetric BB solutions. The most general line element compatible with these symmetries is
\begin{equation}
	ds^{2} = A(r)\,dt^{2} - \frac{dr^{2}}{B(r)} - \Sigma^{2}(r)\bigl(d\theta^{2} + \sin^{2}\theta\,d\phi^{2}\bigr),
	\label{metric}
\end{equation}
where $A(r)$ and $B(r)$ are the metric functions (the lapse and the radial metric component, respectively), and $\Sigma(r)$ is the areal radius. In the standard Simpson--Visser BB, one has $\Sigma(r) = \sqrt{r^{2}+q^{2}}$, where the regularisation parameter $q$ sets the minimum radius at the bounce. In our framework, $\Sigma(r)$ is kept general for the moment, and specific choices will be made when constructing explicit solutions. As we shall see, the 3-form parameter $h_{0}$ will naturally play the role of the bounce scale.

For the metric~\eqref{metric} and the 3-form ansatz~\eqref{H}, the scalar invariants $H_{2}$ and $F_{2}$, respectively, evaluate to
\begin{equation}
	H_{2} = \frac{6\,\chi^{2}(r)}{B(r)},
	\label{H22}
\end{equation}
%
%
\begin{equation}
	\begin{aligned}
		F_{2} = -\frac{6}{A(r)^{2}B(r)^{2}\Sigma^{2}(r)}\,
		\Bigl[
		&A(r)\Sigma(r)\chi(r)B'(r) \\
		&- B(r)\Bigl(4A(r)\chi(r)\Sigma'(r)
		+ \Sigma(r)\bigl(\chi(r)A'(r) + 2A(r)\chi'(r)\bigr)\Bigr)
		\Bigr]^{2}.
	\end{aligned}
	\label{F22}
\end{equation}

Inserting the metric~\eqref{metric} together with the definitions~\eqref{H} and \eqref{F2} into the gravitational field equations~\eqref{eqg}, we obtain three independent components. The $tt$-component, $rr$-component, and $\theta\theta$-component (which equals the $\phi\phi$-component by spherical symmetry) are, respectively,
\begin{equation}
	\begin{aligned}
		&-8\kappa^{2}\mathcal{L}(r)
		-\frac{\kappa^{2}\chi(r)^{2}A'(r)^{2}}{A(r)^{2}}
		+\frac{2\kappa^{2}\chi(r)A'(r)}{A(r)B(r)\Sigma(r)}
		\Bigl[-4B(r)\chi(r)\Sigma'(r)
		+\Sigma(r)\bigl(\chi(r)B'(r)-2B(r)\chi'(r)\bigr)\Bigr] \\
		& -\frac{1}{B(r)^{2}\Sigma(r)^{2}}
		\Bigg[
		\kappa^{2}\Sigma(r)^{2}\chi(r)^{2}B'(r)^{2}
		-4\kappa^{2}B(r)\Sigma(r)\chi(r)
		\Bigl(24\mathcal{L}_{H}(r)\Sigma(r)\chi(r)
		+B'(r)\bigl(2\chi(r)\Sigma'(r)+\Sigma(r)\chi'(r)\bigr)\Bigr) \\
		&+4B(r)^{2}\Bigl(-2+2\kappa^{2}V(r)\Sigma(r)^{2}
		+4\kappa^{2}\chi(r)^{2}\Sigma'(r)^{2}
		+\kappa^{2}\Sigma(r)^{2}\chi'(r)^{2}
		+2\Sigma(r)\Sigma'(r)\bigl(B'(r)+2\kappa^{2}\chi(r)\chi'(r)\bigr)\Bigr) \\
		&+8B(r)^{3}\Bigl(\Sigma'(r)^{2}
		+\Sigma(r)\bigl(\kappa^{2}\epsilon(r)\Sigma(r)\varphi'(r)^{2}
		+2\Sigma''(r)\bigr)\Bigr)
		\Bigg]
		=0,
	\end{aligned}
	\label{eqtt}
\end{equation}
\begin{equation}
	\begin{aligned}
		&-\kappa^{2}B(r)^{2}\Sigma(r)^{2}\chi(r)^{2}A'(r)^{2}
		-2A(r)B(r)\Sigma(r)A'(r)
		\Bigl[4B(r)\bigl(B(r)+\kappa^{2}\chi(r)^{2}\bigr)\Sigma'(r)
		+\kappa^{2}\Sigma(r)\chi(r)\bigl(-\chi(r)B'(r)
		\\
		&
		+2B(r)\chi'(r)\bigr)\Bigr] +A(r)^{2}\Bigg[
		-\kappa^{2}\Sigma(r)^{2}\chi(r)^{2}B'(r)^{2}
		-8B(r)^{3}\Bigl(\Sigma'(r)^{2}
		-\kappa^{2}\epsilon(r)\Sigma(r)^{2}\varphi'(r)^{2}\Bigr) \\
		&+4\kappa^{2}B(r)\Sigma(r)\chi(r)B'(r)
		\Bigl(2\chi(r)\Sigma'(r)+\Sigma(r)\chi'(r)\Bigr)
		\\
		&
		-4B(r)^{2}\Bigg(-2+\kappa^{2}\Bigl(2\mathcal{L}(r)\Sigma(r)^{2}
		+2V(r)\Sigma(r)^{2}
		+\bigl(2\chi(r)\Sigma'(r)+\Sigma(r)\chi'(r)\bigr)^{2}\Bigr)\Bigg)
		\Bigg]
		=0,
	\end{aligned}
	\label{eqrr}
\end{equation}
\begin{equation}
	\begin{aligned}
		&B(r)^{2}\Sigma(r)^{2}\bigl(2B(r)-\kappa^{2}\chi(r)^{2}\bigr)A'(r)^{2}
		+2A(r)B(r)\Sigma(r)\Bigg\{
		-A'(r)\Bigl[-\kappa^{2}\Sigma(r)\chi(r)^{2}B'(r)
		+2B(r)^{2}\Sigma'(r) \\
		&+B(r)\Bigl(4\kappa^{2}\chi(r)^{2}\Sigma'(r)
		+\Sigma(r)\bigl(B'(r)+2\kappa^{2}\chi(r)\chi'(r)\bigr)\Bigr)\Bigr]
		-2B(r)^{2}\Sigma(r)A''(r)\Bigg\} \\
		&-A(r)^{2}\Bigg[
		\kappa^{2}\Sigma(r)^{2}\chi(r)^{2}B'(r)^{2}
		+4B(r)^{2}\Bigl(2\kappa^{2}\mathcal{L}(r)\Sigma(r)^{2}
		+2\kappa^{2}V(r)\Sigma(r)^{2}
		+\Sigma(r)B'(r)\Sigma'(r) \\
		&+4\kappa^{2}\chi(r)^{2}\Sigma'(r)^{2}
		+4\kappa^{2}\Sigma(r)\chi(r)\Sigma'(r)\chi'(r)
		+\kappa^{2}\Sigma(r)^{2}\chi'(r)^{2}\Bigr) \\
		&-4\kappa^{2}B(r)\Sigma(r)\chi(r)
		\Bigl(24\mathcal{L}_{H}(r)\Sigma(r)\chi(r)
		+B'(r)\bigl(2\chi(r)\Sigma'(r)+\Sigma(r)\chi'(r)\bigr)\Bigr) \\
		&+8B(r)^{3}\Sigma(r)\Bigl(\kappa^{2}\epsilon(r)\Sigma(r)\varphi'(r)^{2}
		+\Sigma''(r)\Bigr)
		\Bigg]
		=0,
	\end{aligned}
	\label{eqthetatheta}
\end{equation}
%
where we have introduced the compact notation $\mathcal{L}(r) \equiv \mathcal{L}(H_{2}(r))$, $\mathcal{L}_{H}(r) \equiv d\mathcal{L}/dH_{2}$, and similarly $V(r) \equiv V(\varphi(r))$, $\epsilon(r) \equiv \epsilon(\varphi(r))$.

\subsection{Solving for the matter functions}

The system of equations~\eqref{eqtt}--\eqref{eqthetatheta} determines the metric functions $A(r)$, $B(r)$, $\Sigma(r)$ and the matter functions $\chi(r)$, $\mathcal{L}(r)$, $V(r)$, $\epsilon(r)$, $\varphi(r)$. Since the number of unknown functions exceeds the number of equations, additional physical inputs---such as choosing a specific form for $\Sigma(r)$ or imposing relations among the matter functions---are required to close the system. This is a standard feature of the reconstruction method in BB geometries~\cite{Rodrigues:2023vtm}: one prescribes the geometry (through $\Sigma(r)$) and solves for the matter sources that sustain it.

Following this methodology, we solve the $tt$ and $rr$ components algebraically to express the Lagrangian $\mathcal{L}(r)$ and its derivative $\mathcal{L}_{H}(r)$ in terms of the remaining functions. The resulting expressions are
\begin{equation}
	\begin{aligned}
		\mathcal{L}(r) = -V(r) &+ \frac{1}{8A(r)^{2}B(r)^{2}}
		\Bigg[
		-\chi(r)^{2}\Bigl(B(r)A'(r)-A(r)B'(r)\Bigr)^{2} \\
		&+\frac{4A(r)B(r)}{\kappa^{2}\Sigma(r)^{2}}
		\Bigg(
		-B(r)\Sigma(r)A'(r)\Bigl[2\bigl(B(r)+\kappa^{2}\chi(r)^{2}\bigr)\Sigma'(r)
		+\kappa^{2}\Sigma(r)\chi(r)\chi'(r)\Bigr] \\
		&\qquad +A(r)\Bigl[
		-2B(r)^{2}\Bigl(\Sigma'(r)^{2}-\kappa^{2}\epsilon(r)\Sigma(r)^{2}\varphi'(r)^{2}\Bigr)
		+\kappa^{2}\Sigma(r)\chi(r)B'(r)\Bigl(2\chi(r)\Sigma'(r)+\Sigma(r)\chi'(r)\Bigr) \\
		&\qquad\qquad -B(r)\Bigl(-2+\kappa^{2}\bigl(2\chi(r)\Sigma'(r)+\Sigma(r)\chi'(r)\bigr)^{2}\Bigr)
		\Bigr]
		\Bigg)
		\Bigg],
	\end{aligned}
	\label{L}
\end{equation}
and
\begin{equation}
	\mathcal{L}_{H}(r) = \frac{B(r)}{12\kappa^{2}A(r)\Sigma(r)\chi(r)^{2}}
	\Bigg[
	A(r)B'(r)\Sigma'(r)
	+ B(r)\Bigl(
	-A'(r)\Sigma'(r)
	+ 2A(r)\bigl(\kappa^{2}\epsilon(r)\Sigma(r)\varphi'(r)^{2}
	+\Sigma''(r)\bigr)
	\Bigr)
	\Bigg].
	\label{LH}
\end{equation}

These expressions guarantee that the $tt$ and $rr$ components of the gravitational field equations are identically satisfied, reducing the problem to the $\theta\theta$ component and the matter field equations.

\subsection{The case $B(r)=A(r)$}

A particularly important simplification occurs when we set $B(r)=A(r)$. This choice is natural for static, spherically symmetric solutions with a single metric function, and it encompasses the Schwarzschild and Simpson--Visser families as special cases. Physically, it corresponds to the absence of a radial flux of energy, which is expected for a static configuration in equilibrium. Under this condition, the $tt$ and $rr$ components vanish identically, and the remaining independent $\theta\theta$ component reduces to a remarkably compact form:
\begin{equation}
	-\frac{1}{2}\,A''(r) + \frac{-1 + A(r)\bigl[\Sigma'(r)^{2} + \Sigma(r)\Sigma''(r)\bigr]}{\Sigma(r)^{2}} = 0.
	\label{eq_reduced}
\end{equation}

Equation~\eqref{eq_reduced} is a linear second-order differential equation for $A(r)$ in terms of the prescribed areal radius $\Sigma(r)$. It is one of the central results of this work: for any choice of $\Sigma(r)$, the metric function $A(r)$ is determined by the quadrature
\begin{equation}
	A(r) = \Sigma(r)^{2}\left[\,a_{0} + \int \frac{a_{1} - 2r}{\Sigma(r)^{4}}\;dr\,\right],
	\qquad a_{0},\,a_{1} \in \mathbb{R}.
	\label{A}
\end{equation}
The integration constants $a_{0}$ and $a_{1}$ are determined by appropriate boundary conditions, such as asymptotic flatness ($A(r)\to 1$ as $r\to\infty$) and the requirement that the solution reduces to Schwarzschild--(A)dS in the limit $h_{0}\to 0$. For the Simpson--Visser profile $\Sigma(r)=\sqrt{r^{2}+q^{2}}$, Eq.~\eqref{A} yields the familiar BB metric function.

\subsection{The consistency equation}

An important subtlety arises at this stage. The functions $\mathcal{L}(r)$ and $\mathcal{L}_{H}(r)$ obtained above were derived by solving the $tt$ and $rr$ components of the field equations. However, $\mathcal{L}_{H}(r)$ is, by definition, the derivative of $\mathcal{L}(r)$ with respect to $H_{2}$. This functional relationship imposes the additional constraint
\begin{equation}
	\mathcal{L}_{H}(r) - \frac{d\mathcal{L}(r)}{dr}\left(\frac{dH_{2}(r)}{dr}\right)^{-1} = 0,
	\label{eqconsist}
\end{equation}
which we refer to as the \textit{consistency equation}. Equation~\eqref{eqconsist} ensures that $\mathcal{L}$ is a genuine function of the invariant $H_{2}$, and not merely a radial profile artificially constructed to satisfy the gravitational equations.

The consistency equation can be exploited in two complementary ways, which define the two solution paths explored in this work:
\begin{itemize}
	\item[\textbf{Path A.}] One prescribes the scalar potential $V(r)$ and the scalar field profile $\varphi(r)$, and solves Eq.~\eqref{eqconsist} to determine the kinetic coupling function $\epsilon(r)$. This path leads, for the choice $\epsilon(r)=-1$, to a phantom scalar field and a linear 3-form Lagrangian.
	\item[\textbf{Path B.}] One leaves $\epsilon(r)$ free and instead solves Eq.~\eqref{eqconsist} for the scalar potential $V(r)$. This alternative path yields a constant 3-form Lagrangian and, remarkably, a partially canonical scalar field---a feature that, in the standard NED-based BB framework, is attainable only in modified theories of gravity, but which emerges here within pure general relativity sourced by the self-interacting 3-form field.
\end{itemize}
The following sections are devoted to a detailed exploration of each of these two paths.

\section{Determining the kinetic coupling function $\epsilon(r)$}
\label{sec:epsilon}

In the previous section, we derived the general field equations and the consistency equation~\eqref{eqconsist}, which ensures that the Lagrangian $\mathcal{L}(H_2)$ is a genuine function of the 3-form invariant $H_2$. In this section, we follow \textbf{Path A}: we solve the consistency equation to determine the kinetic coupling function $\epsilon(r)$, given suitable choices for the scalar potential $V(r)$ and the scalar field profile $\varphi(r)$. This path leads to a solution with a phantom scalar field, $\epsilon(r)=-1$, which is the typical configuration for supporting wormhole throats and bouncing geometries in general relativity.

\subsection{General solution of the consistency equation}

Imposing the consistency condition~\eqref{eqconsist}, together with the expressions~\eqref{L} and~\eqref{LH} for $\mathcal{L}(r)$ and $\mathcal{L}_{H}(r)$, yields a differential equation for $\epsilon(r)$. After a lengthy but straightforward calculation, the general solution can be cast in the following compact form:
\begin{equation}
	\begin{aligned}
		\epsilon(r) =& \frac{\chi(r)^{2}}{A(r)^{2}\varphi'(r)^{2}}
		\int \frac{A(r)\,dr}{\kappa^{2}\Sigma(r)^{3}\chi(r)^{3}}
		\Bigg[
		2\chi(r)\Sigma'(r)\Bigl(1 - \bigl(A(r)+2\kappa^{2}\chi(r)^{2}\bigr)\Sigma'(r)^{2}\Bigr) \\
		&+ 2\Sigma(r)\chi(r)\Sigma'(r)\Bigl(\kappa^{2}\chi(r)\Sigma'(r)\chi'(r) + \bigl(A(r)+2\kappa^{2}\chi(r)^{2}\bigr)\Sigma''(r)\Bigr) 
		+ \kappa^{2}\Sigma(r)^{3}\chi(r)\Bigl(V'(r) + \chi'(r)\chi''(r)\Bigr) \\
		&+ \Sigma(r)^{2}\Bigg[\chi(r)\Sigma'(r)\Bigl(2\kappa^{2}\chi'(r)^{2} + A''(r)\Bigr) 
		 + 2A(r)\chi'(r)\Sigma''(r) + 2\kappa^{2}\chi(r)^{2}\Bigl(\chi'(r)\Sigma''(r) + \Sigma'(r)\chi''(r)\Bigr)\Bigg]\Bigg]\,dr.
	\end{aligned}
	\label{epsilon_general}
\end{equation}
This expression is exact and fully determines $\epsilon(r)$ once the functions $A(r)$, $\Sigma(r)$, $\chi(r)$, $V(r)$, and $\varphi(r)$ are specified. The complexity of this formula reflects the non-linear interplay between the 3-form field, the scalar field, and the spacetime geometry. However, as we shall see, it simplifies dramatically when the matter field equations are taken into account.

\subsection{The 3-form profile and the scalar field equation}

A crucial simplification occurs when we examine the scalar field equation of motion~\eqref{eqphi}. Requiring that this equation be satisfied imposes a simple and elegant relation between the 3-form function $\chi(r)$ and the areal radius $\Sigma(r)$:
\begin{equation}
	\chi(r) = \frac{h_{0}}{\Sigma^{2}(r)}.
	\label{chi}
\end{equation}
The integration constant $h_{0}$ is the parameter that governs the strength of the 3-form field $H_{abc}$. When $h_{0} \neq 0$, the 3-form contributes as a non-trivial matter source that curves spacetime, and its energy density is distributed according to the inverse fourth power of the areal radius. In the limit $h_{0} \to 0$, the 3-form field vanishes, and we recover vacuum solutions of general relativity. As we shall see, $h_{0}$ also plays the role of the bounce parameter in the BB geometry, in complete analogy with the electric or magnetic charge in NED-based BB solutions.

Substituting the profile~\eqref{chi} into Eq.~\eqref{epsilon_general}, the expression for $\epsilon(r)$ simplifies dramatically. After integration by parts and using the reduced field equation~\eqref{eq_reduced}, we obtain
\begin{equation}
	\epsilon(r) = \frac{\displaystyle\int A(r)\,\Sigma(r)^{4}\,V'(r)\,dr}{A(r)^{2}\,\Sigma(r)^{4}\,\varphi'(r)^{2}}.
	\label{epsilon_simplified}
\end{equation}
This compact formula shows that the kinetic coupling function is determined by the interplay between the scalar potential $V(r)$, the metric function $A(r)$, and the scalar field profile $\varphi(r)$. The denominator, proportional to $A(r)^{2}\varphi'(r)^{2}$, indicates that $\epsilon(r)$ can change sign depending on the behavior of these functions, which is intimately related to the canonical or phantom nature of the scalar field.

\subsection{Fixing $\epsilon(r) = -1$: the phantom scalar field}

In the standard Simpson--Visser BB framework~\cite{Rodrigues:2023vtm}, the scalar field is of the phantom type, characterized by $\epsilon(r) = -1$. This choice is physically well-motivated: phantom fields possess a negative kinetic energy, which allows them to violate the null energy condition in a controlled manner and thereby sustain wormhole throats and bouncing geometries without introducing exotic matter in an ad hoc fashion. Imposing $\epsilon(r) = -1$ in Eq.~\eqref{epsilon_simplified} yields a first-order differential equation for the scalar potential $V(r)$:
\begin{equation}
	V'(r) = -\frac{\dfrac{d}{dr}\Bigl[A(r)^{2}\,\Sigma(r)^{4}\,\varphi'(r)^{2}\Bigr]}{A(r)\,\Sigma(r)^{4}}.
\end{equation}
Integrating this equation, we obtain the scalar potential in closed form,
\begin{equation}
	V(r) = -\int \frac{\dfrac{d}{dr}\Bigl[A(r)^{2}\,\Sigma(r)^{4}\,\varphi'(r)^{2}\Bigr]}{A(r)\,\Sigma(r)^{4}}\;dr.
	\label{V1}
\end{equation}
Equation~\eqref{V1} determines the scalar potential $V(r)$ self-consistently, once $A(r)$, $\Sigma(r)$, and $\varphi(r)$ are known. This is a standard result of the reconstruction method: the geometry dictates the matter content required to support it.

\subsection{The black-bounce profile and the metric function}

Following the Simpson--Visser construction, we identify the 3-form parameter $h_{0}$ as the bounce scale and choose the areal radius
\begin{equation}
	\Sigma(r) = \sqrt{r^{2} + h_{0}^{2}}.
	\label{Sigma1}
\end{equation}
This choice ensures that the geometry is regular at $r=0$, where $\Sigma(0)=h_{0}>0$, and that the two asymptotic regions $r\to+\infty$ and $r\to-\infty$ are connected by a throat of minimum radius $h_{0}$. The bounce is thus directly controlled by the 3-form field strength: the larger the 3-form parameter, the larger the minimum radius and the more pronounced the deviation from the Schwarzschild geometry.

The integration constants $a_{0}$ and $a_{1}$ in the general solution~\eqref{A} are fixed by requiring the correct Schwarzschild--(anti-)de Sitter asymptotic behavior. We set
\begin{equation}
	a_{0} = -\frac{1}{6}\left(\frac{9\pi M}{h_{0}^{3}} + 2\Lambda\right),\qquad
	a_{1} = 6M,
\end{equation}
where $M$ is the mass parameter and $\Lambda$ is the cosmological constant. Substituting these constants and the profile~\eqref{Sigma1} into Eq.~\eqref{A}, we obtain the metric function
\begin{equation}
	\begin{aligned}
		A(r) = 1 - \frac{3\pi M}{2h_{0}} + \frac{3Mr}{h_{0}^{2}} - \frac{3\pi Mr^{2}}{2h_{0}^{3}}
		- \frac{h_{0}^{2}\Lambda}{3} - \frac{r^{2}\Lambda}{3} + \frac{3M}{h_{0}}\arctan\!\left(\frac{r}{h_{0}}\right)
		+ \frac{3Mr^{2}}{h_{0}^{3}}\arctan\!\left(\frac{r}{h_{0}}\right).
	\end{aligned}
	\label{a1}
\end{equation}
The presence of the arctangent terms is a distinctive signature of the 3-form field: it reflects the non-trivial distribution of the 3-form energy density along the radial direction, which deviates from the simple $1/r$ fall-off characteristic of the Schwarzschild solution.

The scalar field profile consistent with the 3-form equation of motion~\eqref{eqH} is found to be
\begin{equation}
	\varphi(r) = \frac{1}{\kappa}\arctan\!\left(\frac{r}{h_{0}}\right).
	\label{varphi}
\end{equation}
Since the radial coordinate ranges over $r \in (-\infty,\infty)$, the scalar field is confined to the finite interval $\varphi \in [-\pi/(2\kappa),\,\pi/(2\kappa)]$. This bounded nature of the scalar field is a characteristic feature of BB solutions and ensures that the scalar potential $V(\varphi)$ remains well-defined and free of singularities over the entire domain of the field.

\subsection{Properties of the solution}

\subsubsection{Limiting cases and asymptotic behavior}

If the 3-form field is switched off by setting $h_{0}=0$, the metric function~\eqref{a1} reduces, for the positive radial branch $r \ge 0$, to the familiar Schwarzschild--(anti-)de Sitter solution:
\begin{equation}
	A(r) = 1 - \frac{2M}{r} - \frac{\Lambda}{3}\,r^{2}.
\end{equation}
Thus, our solution contains the standard $\Lambda$-vacuum black hole as a limiting case, with the 3-form parameter $h_{0}$ controlling the deviation from general relativity.

The asymptotic behavior of the metric function at large positive and negative $r$, and near the bounce at $r=0$, is given by
\begin{align}
	A(r \to +\infty) &\sim 1 - \frac{2M}{r} - \frac{h_{0}^{2}\Lambda}{3} - \frac{r^{2}\Lambda}{3},\\[4pt]
	A(r \to -\infty) &\sim 1 - \frac{3\pi M}{h_{0}} - \frac{2M}{r} - \frac{3\pi Mr^{2}}{h_{0}^{3}} - \frac{h_{0}^{2}\Lambda}{3} - \frac{r^{2}\Lambda}{3},\\[4pt]
	A(r \to 0) &\sim 1 - \frac{3\pi M}{2h_{0}} - \frac{h_{0}^{2}\Lambda}{3} + \frac{6M}{h_{0}^{2}}\,r
	- \left(\frac{3\pi M}{2h_{0}^{3}} + \frac{\Lambda}{3}\right)r^{2} + \mathcal{O}(r^{3}).
\end{align}

The solution is asymptotically de Sitter (dS) for $\Lambda>0$ and anti-de Sitter (AdS) for $\Lambda<0$ in the $r>0$ branch. The $r<0$ branch exhibits a more complex asymptotic structure due to the polynomial terms generated by the arctangent, reflecting the asymmetric nature of the 3-form source. Near $r=0$, the metric is regular, with $A(0) = 1 - \frac{3\pi M}{2h_{0}} - \frac{h_{0}^{2}\Lambda}{3}$, which may be positive or negative depending on the parameter values.

\begin{figure}[ht!]
	\centering
	\includegraphics[width=0.495\textwidth]{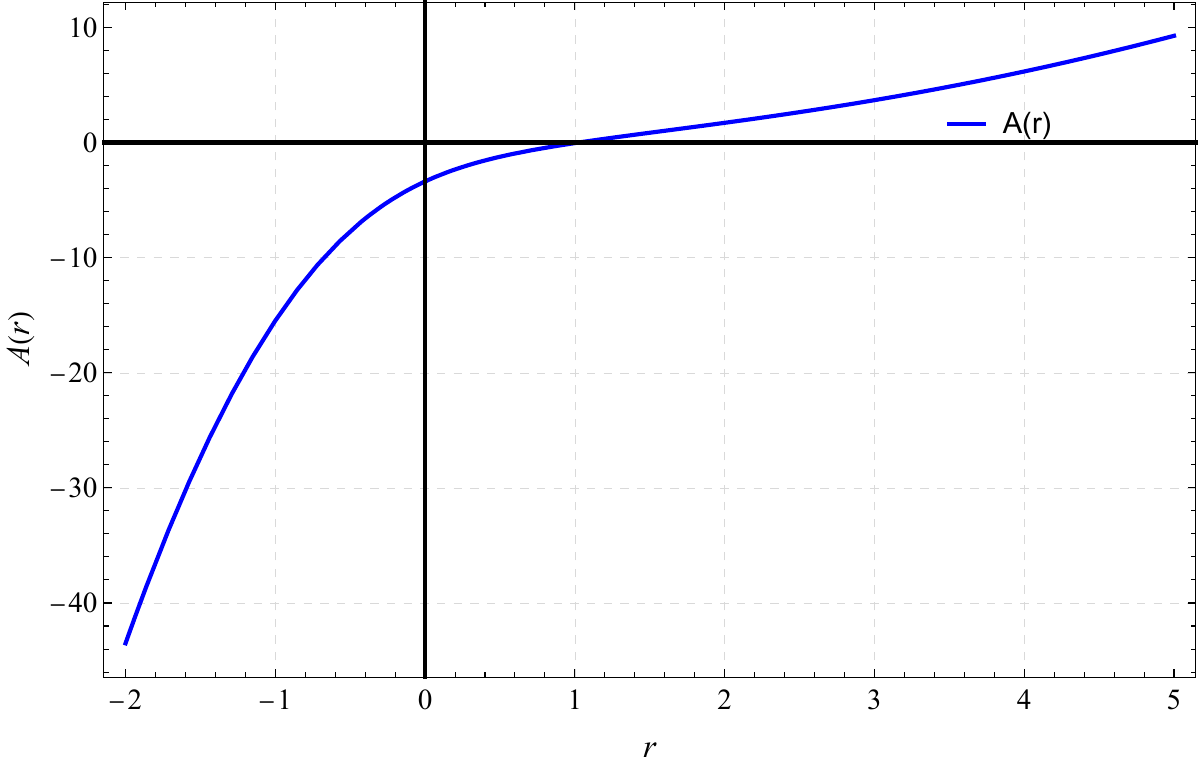}
	\includegraphics[width=0.495\textwidth]{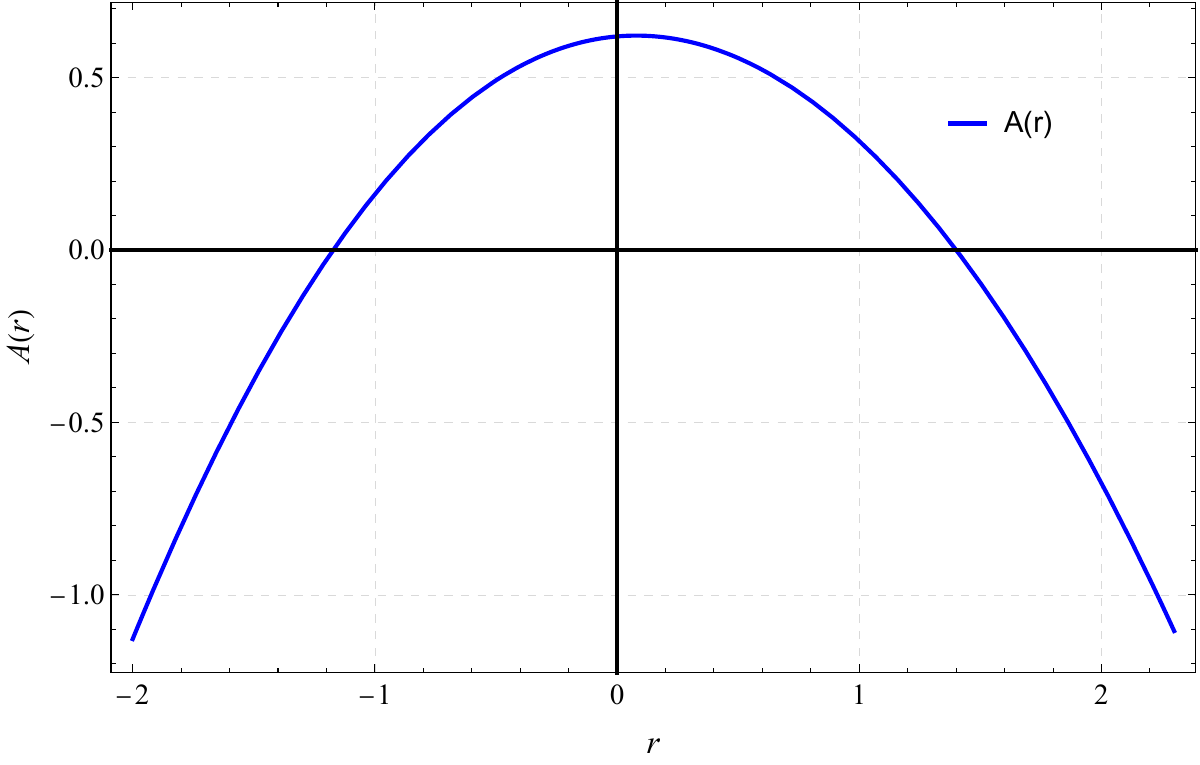}
	\caption{Graphical representation of the metric function $A(r)$ given by Eq.~\eqref{a1}. \textbf{Left panel:} AdS case with parameters $\{M=h_{0}=1,\;\Lambda=-1\}$. \textbf{Right panel:} dS case with parameters $\{M=0.01,\;h_{0}=\Lambda=1\}$.}
	\label{A-epsilon}
\end{figure}

The solution is asymmetric: the positive and negative $r$ branches possess distinct horizon structures. In the AdS case ($\Lambda<0$), there is typically a single event horizon, as illustrated in the left panel of Fig.~\ref{A-epsilon}. In the dS case ($\Lambda>0$), two cosmological horizons may appear---one in each branch---as shown in the right panel of Fig.~\ref{A-epsilon}. This asymmetry is a direct consequence of the 3-form field configuration and distinguishes these solutions from the symmetric Simpson--Visser BB.

\subsubsection{The matter sector}

For the solution~\eqref{a1}, the 3-form invariant $H_{2}$ and the Lagrangian are given by
\begin{eqnarray}
		H_{2} = -\frac{36h_{0}^{5}}{(h_{0}^{2}+r^{2})^{2}}\Bigg[
		9h_{0}M\Bigl(-2r + \frac{\pi}{h_{0}}(h_{0}^{2}+r^{2})\Bigr) 
		+ 2h_{0}^{5}\Lambda
			 + 2h_{0}^{3}(-3+r^{2}\Lambda)
		- 18M(h_{0}^{2}+r^{2})\arctan\!\left(\frac{r}{h_{0}}\right)\Bigg]^{-1},
\end{eqnarray}
\begin{equation}
	\mathcal{L}_{H} = \frac{1}{216\,\kappa^{2}}\left[ \dfrac{81M^{2}\pi^{2}}{h_{0}^{2}} + \dfrac{36\pi M}{h_{0}}\bigl(-3+h_{0}^{2}\Lambda\bigr) + 4\bigl(-3+h_{0}^{2}\Lambda\bigr)^{2} \right],
\end{equation}
\begin{equation}
	\mathcal{L}(H_{2}) = \mathcal{L}_{H}\,H_{2} + \frac{1}{2\kappa^{2}}\left( \dfrac{9\pi M}{h_{0}^{3}} + 2\Lambda\right).
\end{equation}

Remarkably, the Lagrangian $\mathcal{L}(H_{2})$ is \textit{linear} in $H_{2}$ for this solution, meaning that the self-interaction reduces to a simple mass-like term for the 3-form field. The constant offset in $\mathcal{L}(H_{2})$ acts as an effective cosmological constant contribution, which, together with the explicit $\Lambda$ term, determines the asymptotic behavior of the spacetime.

Inverting the relation $r = h_{0}\tan(\kappa\varphi)$ from Eq.~\eqref{varphi} and substituting into Eq.~\eqref{V1}, the scalar potential expressed in terms of the field $\varphi$ reads
\begin{equation}
		V(\varphi) = -\frac{\cos^{2}(\kappa\varphi)}{3h_{0}^{3}\kappa^{2}}
		\left[
		\frac{9\pi M}{h_{0}^{2}} + 2h_{0}^{3}\Lambda + 27M\tan(\kappa\varphi) 
		+ 9M\kappa\varphi\Bigl(1 + 3\tan^{2}(\kappa\varphi)\Bigr)\right].
	\label{V2}
\end{equation}

\begin{figure}[ht!]
	\centering
	\includegraphics[width=0.495\textwidth]{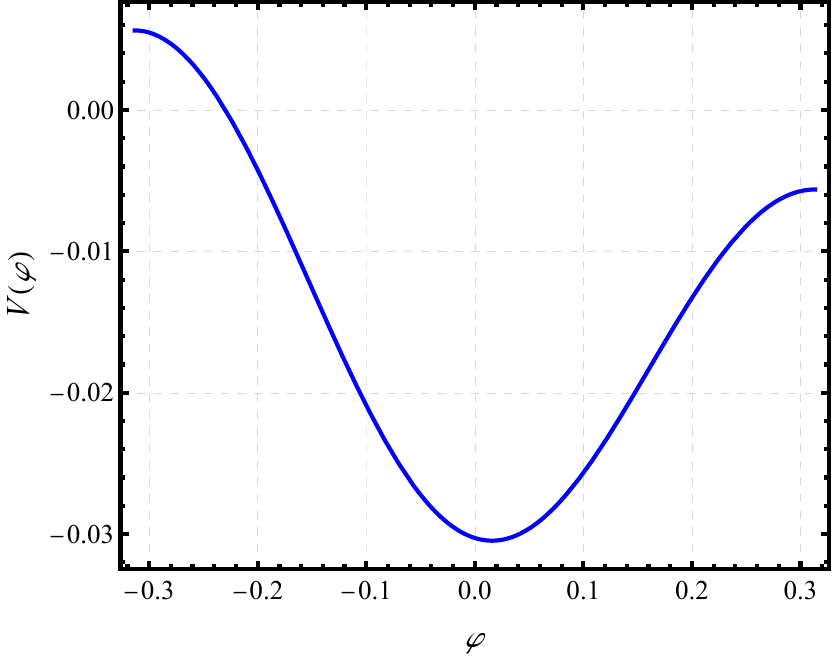}
	\caption{Graphical representation of the scalar potential $V(\varphi)$ given by Eq.~\eqref{V2}, for the parameters $\{M=0.01,\;h_{0}=\Lambda=1\}$.}
	\label{V-epsilon}
\end{figure}

Figure~\ref{V-epsilon} displays the potential $V(\varphi)$ for a representative set of parameters. The potential is bounded and well-defined over the entire range of the scalar field, ensuring that the matter sector is free of pathologies. The oscillatory behavior visible in the plot reflects the interplay between the polynomial and trigonometric terms in Eq.~\eqref{V2}.

\subsubsection{Regularity}

The regularity of the solution is confirmed by computing the Kretschmann scalar $K = R_{abcd}R^{abcd}$. An exact, albeit lengthy, expression for $K$ can be obtained using algebraic computing software. We find that $K$ is finite everywhere, including at the bounce $r=0$ and in the asymptotic limits $r\to\pm\infty$, where it approaches the (A)dS value $K \to 8\Lambda^{2}/3$. The absence of curvature singularities confirms that the solution describes a globally regular, asymmetric black-bounce geometry sourced by a self-interacting 3-form field and a phantom scalar field. This solution interpolates smoothly between Schwarzschild--(A)dS configurations in the limit $h_{0}\to 0$ and regular BB configurations for $h_{0}\neq 0$, thereby providing a consistent and physically well-motivated extension of the standard black-hole paradigm.


\section{Determining the scalar potential $V(r)$}
\label{sec:potential}

In the previous section, we followed \textbf{Path A} and solved the consistency equation~\eqref{eqconsist} to determine the kinetic coupling function $\epsilon(r)$, fixing $\epsilon(r)=-1$ to obtain a phantom scalar field. In this section, we follow the complementary \textbf{Path B}: we solve the consistency equation to determine the scalar potential $V(r)$, leaving $\epsilon(r)$ free to be determined by the field equations. This approach yields a new class of BB solutions with a partially canonical scalar field---a feature that, in the standard NED-based framework, arises only in modified theories of gravity~\cite{deSousaSilva:2026kvp,Silva:2025fqj}, but which we obtain here within pure general relativity sourced by a self-interacting 3-form field. The existence of two distinct solution paths highlights the richness of the 3-form matter sector and its ability to accommodate different physical configurations depending on how the consistency condition is resolved.

\subsection{General solution for $V(r)$}

Solving the consistency equation~\eqref{eqconsist} for $V(r)$, rather than for $\epsilon(r)$, yields the following integral expression for the scalar potential:
\begin{equation}
	\begin{aligned}
		V(r) &= \int \frac{1}{\kappa^{2}\Sigma(r)^{3}\chi(r)}\,
		\Bigg[ -2A(r)\Sigma(r)^{2}\chi'(r)\Bigl(\kappa^{2}\epsilon(r)\Sigma(r)\varphi'(r)^{2} + \Sigma''(r)\Bigr) \\
		&\qquad + 4\kappa^{2}\chi(r)^{3}\Sigma'(r)\Bigl(\Sigma'(r)^{2} - \Sigma(r)\Sigma''(r)\Bigr) \\
		&\qquad - 2\kappa^{2}\Sigma(r)\chi(r)^{2}\Bigl(\chi'(r)\bigl(\Sigma'(r)^{2} + \Sigma(r)\Sigma''(r)\bigr) + \Sigma(r)\Sigma'(r)\chi''(r)\Bigr) \\
		&\qquad + \chi(r)\Bigg(2A(r)\Sigma'(r)^{3} - \Sigma'(r)\Bigl[2 + \Sigma(r)^{2}\bigl(2\kappa^{2}\chi'(r)^{2} + A''(r)\bigr) + 2A(r)\Sigma(r)\Sigma''(r)\Bigr] \\
		&\qquad + \kappa^{2}\Sigma(r)^{3}\Bigl(\varphi'(r)\bigl[(2\epsilon(r)A'(r) + A(r)\epsilon'(r))\varphi'(r) + 2A(r)\epsilon(r)\varphi''(r)\bigr] - \chi'(r)\chi''(r)\Bigr)\Bigg)\Bigg]\,dr.
	\end{aligned}
	\label{V_general}
\end{equation}
This expression, while considerably more involved than its counterpart in Path A, is exact and completely general. It determines the scalar potential $V(r)$ once the metric functions $A(r)$, $\Sigma(r)$, the 3-form profile $\chi(r)$, the kinetic coupling $\epsilon(r)$, and the scalar field $\varphi(r)$ are specified. The complexity of this formula reflects the fact that, in Path B, the scalar potential must compensate for the freedom in $\epsilon(r)$ in order to satisfy the full set of field equations.

\subsection{Constraints from the matter field equations}

As in the previous section, the scalar field equation of motion~\eqref{eqphi} forces the 3-form profile to take the same simple form,
\begin{equation}
	\chi(r) = \frac{h_{0}}{\Sigma^{2}(r)},
	\label{chi2}
\end{equation}
where $h_{0}$ is again the 3-form coupling constant. This universality of the 3-form profile—arising independently of whether one follows Path A or Path B—underscores the fundamental role played by the 3-form field in determining the bounce scale. Substituting this profile into the 3-form field equation~\eqref{eqH} and solving for the kinetic coupling function yields a remarkably compact result:
\begin{equation}
	\epsilon(r) = -\frac{\Sigma''(r)}{\kappa^{2}\,\Sigma(r)\,\varphi'(r)^{2}}.
	\label{epsilon_from_H}
\end{equation}

Equation~\eqref{epsilon_from_H} is identical in form to the expression obtained in the standard NED-based BB framework~\cite{Rodrigues:2023vtm}. It reveals a profound geometric connection: the nature of the scalar field---whether it is canonical ($\epsilon>0$) or phantom ($\epsilon<0$)---is determined solely by the sign of the second derivative of the areal radius, $\Sigma''(r)$. For the usual Simpson--Visser profile $\Sigma(r)=\sqrt{r^{2}+h_{0}^{2}}$, one finds $\Sigma''(r)>0$ everywhere, leading to $\epsilon(r)<0$ and therefore a globally phantom scalar field. To obtain a canonical scalar field outside the event horizon, we must deform the areal radius in such a way that $\Sigma''(r)$ changes sign, becoming negative in the exterior region while remaining positive near the bounce.

\subsection{A modified areal radius for a canonical scalar field}

Guided by the condition $\epsilon(r)>0$ in the exterior region, we introduce a deformation of the Simpson--Visser profile by adding a rational correction term that softens the behavior at large $r$:
\begin{equation}
	\Sigma(r) = \sqrt{r^{2}+h_{0}^{2}} - \frac{k_{1}h_{0}^{2}}{\sqrt{r^{2}+h_{0}^{2}}},\qquad 0 \le k_{1} < 1.
	\label{Sigma2}
\end{equation}

The dimensionless parameter $k_{1}$ controls the deviation from the pure Simpson--Visser profile, which is recovered for $k_{1}=0$. The condition $k_{1}<1$ ensures that $\Sigma(r)$ remains real and positive for all $r$, preserving the regularity of the geometry. The physical interpretation of this deformation is that the 3-form field, through its coupling to the scalar sector, effectively modifies the shape of the throat, stretching it in the radial direction. The second derivative of this profile changes sign at a certain radius, allowing $\epsilon(r)$ to become positive in the exterior region while remaining negative near the bounce---thereby yielding a \textit{partially canonical} scalar field that is phantom only in the interior region where exotic matter is required to sustain the bounce.

\subsection{The metric function}

With the areal radius~\eqref{Sigma2} and the 3-form profile~\eqref{chi2}, the general solution~\eqref{A} for the metric function becomes
\begin{equation}
	\begin{aligned}
		A(r) = \frac{\bigl(-h_{0}^{2}(-1+k_{1})+r^{2}\bigr)^{2}}{48\,(h_{0}^{2}+r^{2})}\,
		\Bigg[
		&-\frac{9\pi M\bigl(8+(-4+k_{1})k_{1}\bigr)}{(-1+k_{1})^{2}\left(-\dfrac{1}{h_{0}^{2}(-1+k_{1})}\right)^{3/2}}
		- 16\Lambda \\
		&+\frac{1}{h_{0}^{3}}\Bigg(
		-\frac{2h_{0}}{(-1+k_{1})^{3}\bigl(-h_{0}^{2}(-1+k_{1})+r^{2}\bigr)^{3}} \\
		&\qquad\times \Bigl[-8h_{0}^{6}(-1+k_{1})^{3}\bigl(3+(-3+k_{1})k_{1}\bigr) + 9M\bigl(8+(-4+k_{1})k_{1}\bigr)r^{5} \\
		&\qquad\quad + 3h_{0}^{4}(-1+k_{1})^{2}r\Bigl(-3M(-8+(-4+k_{1})k_{1}) + 8(-2+k_{1})(-1+k_{1})r\Bigr) \\
		&\qquad\quad - 24h_{0}^{2}(-1+k_{1})r^{3}\Bigl(-M(-6+k_{1}^{2}) + (-1+k_{1})^{2}r\Bigr)\Bigr] \\
		&+ \frac{18M\bigl(8+(-4+k_{1})k_{1}\bigr)\,\operatorname{arccoth}\!\left(\dfrac{h_{0}\sqrt{-1+k_{1}}}{r}\right)}{(-1+k_{1})^{7/2}}
		\Bigg)\Bigg],
	\end{aligned}
	\label{a2}
\end{equation}
with the integration constants calibrated as
\begin{equation}
	a_{0} = -\frac{3\pi M\bigl(8+(-4+k_{1})k_{1}\bigr)}{16(-1+k_{1})^{2}\left(-\dfrac{1}{h_{0}^{2}(-1+k_{1})}\right)^{3/2}} - \frac{\Lambda}{3},\qquad \qquad
	a_{1} = 6M.
\end{equation}
The presence of the inverse hyperbolic cotangent function, $\operatorname{arccoth}$, signals the distinct asymptotic structure of this solution compared to the arctangent-based form of Path A. This difference originates from the deformation parameter $k_{1}$, which modifies the algebraic structure of the metric function through the rational correction in $\Sigma(r)$.

\subsection{Properties of the solution}

\subsubsection{Horizon structure and asymptotics}

\begin{figure}[ht!]
	\centering
	\includegraphics[width=0.495\textwidth]{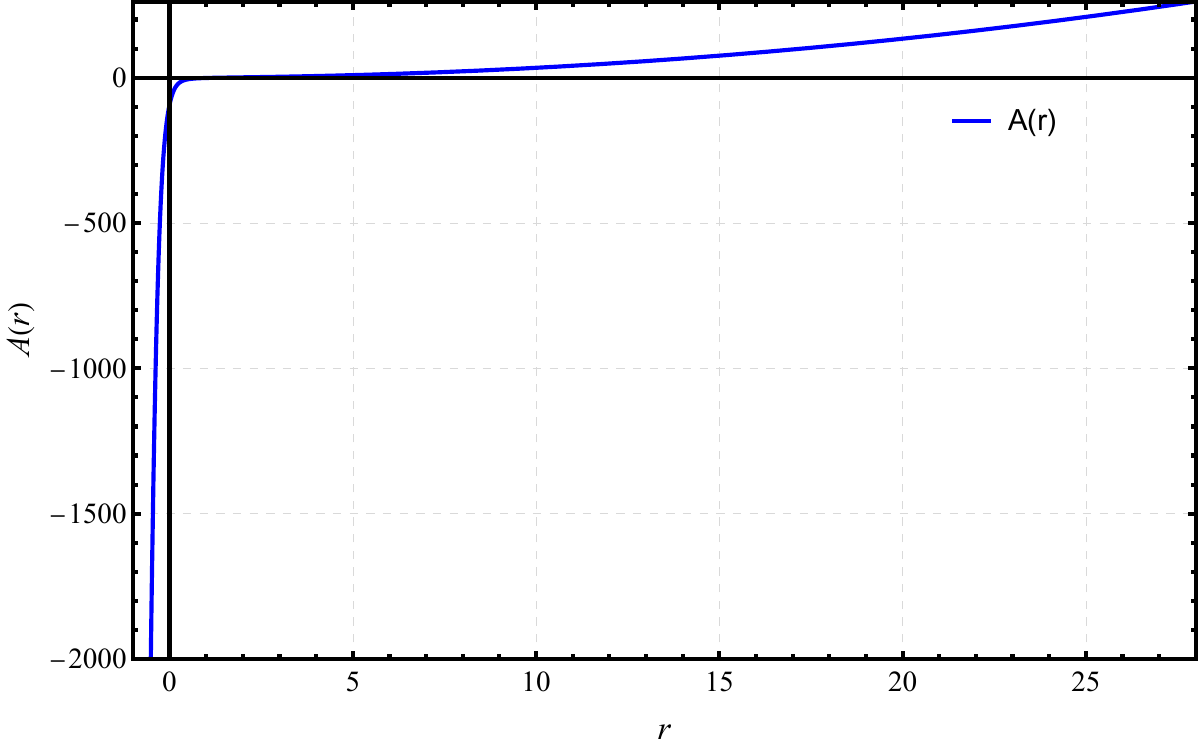}
	\includegraphics[width=0.495\textwidth]{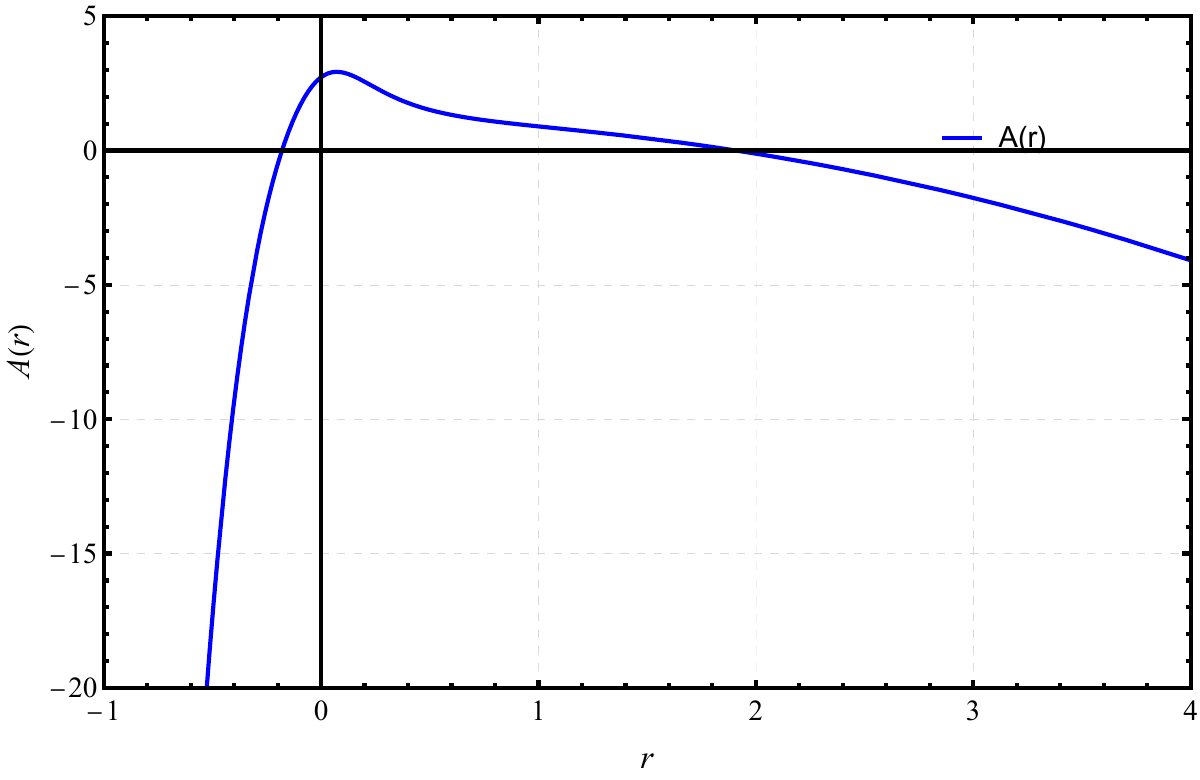}
	\caption{Graphical representation of the metric function $A(r)$ given by Eq.~\eqref{a2}. \textbf{Left panel:} AdS case with parameters $\{k_{1}=0.9,\;\Lambda=-1,\;h_{0}=M=1\}$. \textbf{Right panel:} dS case with parameters $\{k_{1}=0.9,\;\Lambda=h_{0}=1,\;M=0.01\}$.}
	\label{A-V}
\end{figure}

The metric function~\eqref{a2} shares the same qualitative features as the first solution~\eqref{a1}: it is asymmetric, with distinct horizon structures in the positive and negative $r$ branches, and reduces to Schwarzschild--(A)dS for appropriate parameter limits. The deformation parameter $k_{1}$ provides an additional degree of freedom that allows the horizon positions and the asymptotic behavior to be tuned independently of the bounce scale $h_{0}$. Figure~\ref{A-V} displays the behavior of $A(r)$ for representative AdS and dS configurations, illustrating how the horizon structure responds to variations in the parameters.

\subsubsection{Regularity}

\begin{figure}[ht!]
	\centering
	\includegraphics[width=0.495\textwidth]{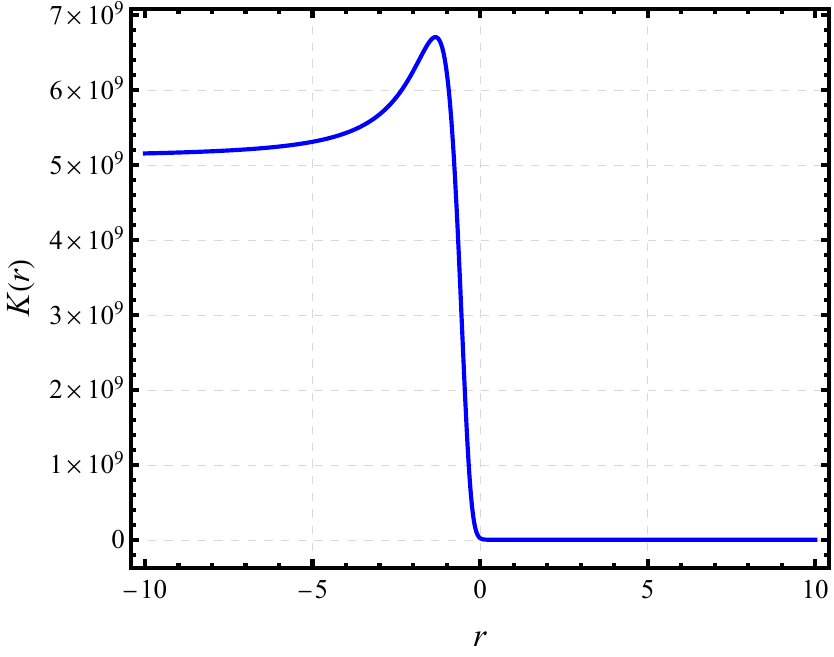}
	\caption{Graphical representation of the Kretschmann scalar $K(r)$ for the solution~\eqref{a2}--\eqref{Sigma2}, with parameters $\{M=1,\;h_{0}=1.1,\;\Lambda=-1,\;k_{1}=0.9\}$.}
	\label{K1}
\end{figure}

The Kretschmann scalar $K = R_{abcd}R^{abcd}$ for this solution has been computed analytically and is found to be finite everywhere, including at the bounce $r=0$ and in the asymptotic limits $r\to\pm\infty$. Figure~\ref{K1} illustrates the radial profile of $K$ for a representative choice of parameters. The smooth, peaked structure near $r=0$ reflects the curvature concentration at the bounce, while the rapid fall-off at large $|r|$ confirms the asymptotic (A)dS behavior. The global regularity of $K$ provides a robust verification that the solution is free of curvature singularities.

\subsubsection{The matter sector}

A striking feature of this second solution is that the 3-form Lagrangian $\mathcal{L}(H_{2})$ is a \textit{constant}:
\begin{equation}
	\mathcal{L}(H_{2}) = \frac{9\pi M\bigl(8-4k_{1}+k_{1}^{2}\bigr)}{16(-1+k_{1})^{2}\kappa^{2}\left(-\dfrac{1}{h_{0}^{2}(-1+k_{1})}\right)^{3/2}} + \frac{\Lambda}{\kappa^{2}},
	\label{L_constant}
\end{equation}
and consequently its derivative vanishes, $\mathcal{L}_{H} \equiv 0$. This means that the self-interaction term in the action reduces to an effective cosmological constant, and the 3-form field contributes to the geometry solely through its kinetic term $F_{2}$. The constant value of $\mathcal{L}(H_{2})$ depends on the mass $M$, the bounce scale $h_{0}$, the deformation parameter $k_{1}$, and the cosmological constant $\Lambda$, intertwining the 3-form sector with the global properties of the spacetime in a non-trivial manner.

\subsubsection{The scalar field: a partially canonical nature}

\begin{figure}[th!]
	\centering
	\includegraphics[width=0.495\textwidth]{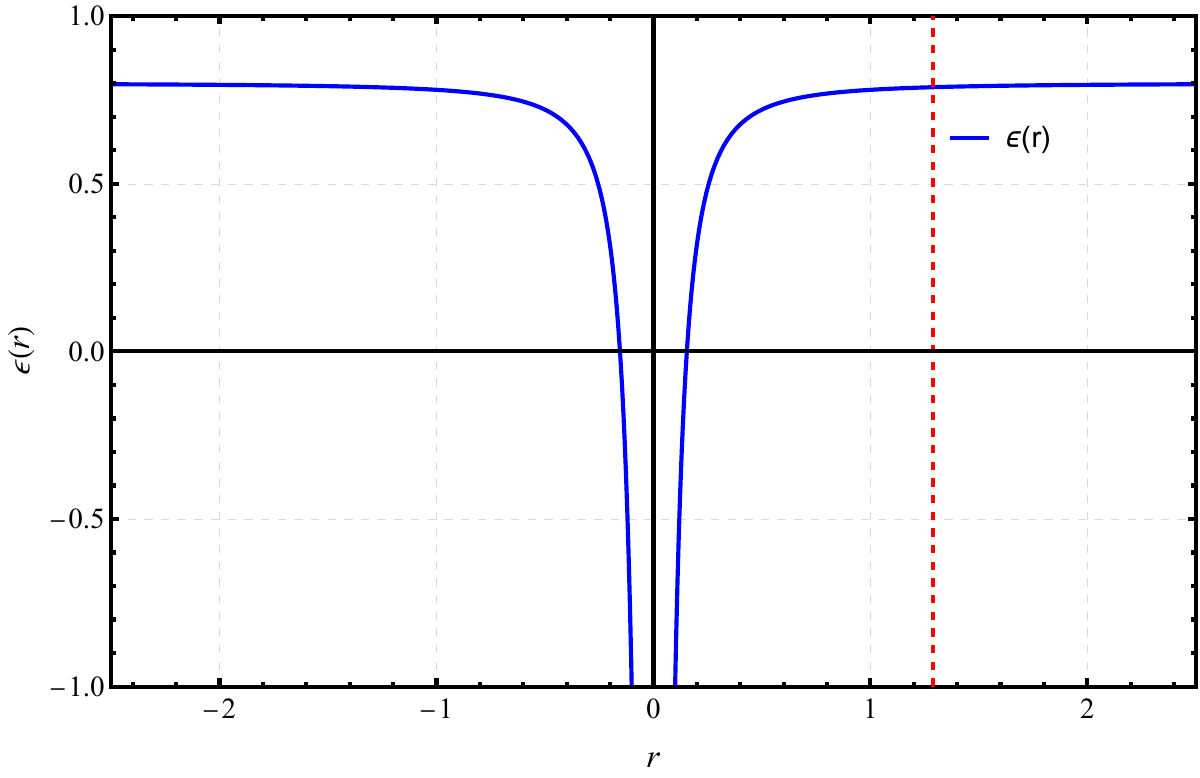}
	\caption{Graphical representation of the kinetic coupling function $\epsilon(r)$ for the solution~\eqref{a2}--\eqref{Sigma2}, with parameters $\{h_{0}=0.1,\;k_{1}=0.9\}$. The red dashed vertical line marks the position of the event horizon at $r_{H}=1.28992$ (for $M=1$, $\Lambda=-1$). Note that $\epsilon(r)>0$ for $r>r_{H}$, indicating a canonical scalar field in the exterior region.}
	\label{epsilon}
\end{figure}

The scalar field profile is again given by the arctangent form, $\varphi(r) = \frac{1}{\kappa}\arctan(r/h_{0})$, which is universal to both solution paths. The kinetic coupling function $\epsilon(r)$, determined by Eq.~\eqref{epsilon_from_H}, is plotted in Fig.~\ref{epsilon}. The key result is that $\epsilon(r) > 0$ in the exterior region $r > r_{H}$ (where $r_{H}$ denotes the event horizon radius), indicating a \textit{canonical} scalar field outside the black hole. Near the bounce, $\epsilon(r)$ becomes negative, signalling a transition to a phantom behavior in the interior region where the bounce is located.

This behavior is particularly noteworthy and constitutes one of the main results of this work. In the standard BB framework sourced by NED and a scalar field, the scalar field is necessarily phantom ($\epsilon<0$) throughout the entire spacetime when working within pure general relativity. Canonical scalar fields supporting BB solutions have previously been obtained only in modified theories of gravity, such as $f(R)$ gravity~\cite{deSousaSilva:2026kvp,Silva:2025fqj}. Here, we achieve a partially canonical scalar field within \textit{general relativity} by virtue of the self-interacting 3-form matter sector, which provides the additional degrees of freedom necessary to support this configuration. The transition from phantom to canonical behavior at the horizon is a direct consequence of the sign change in $\Sigma''(r)$ induced by the deformation parameter $k_{1}$.

\subsection{Summary of the two solution paths}

The two paths explored in this and the previous section illustrate the versatility of the 3-form framework for constructing regular BB solutions. In Path A (Section~\ref{sec:epsilon}), fixing $\epsilon(r)=-1$ to obtain a globally phantom scalar field leads to a linear 3-form Lagrangian $\mathcal{L}(H_{2})$ and a solution that reduces to Schwarzschild--(A)dS in the limit $h_{0}\to 0$, with the standard Simpson--Visser areal radius $\Sigma(r)=\sqrt{r^{2}+h_{0}^{2}}$. In Path B (this section), leaving $\epsilon(r)$ free and solving for $V(r)$ yields a constant 3-form Lagrangian and a solution with a partially canonical scalar field, where the areal radius is deformed by the additional parameter $k_{1}$ that controls the transition between canonical and phantom behavior. Both solutions are globally regular, asymptotically (A)dS (or flat for $\Lambda=0$), and exhibit the characteristic asymmetric horizon structure of BB geometries. The existence of these two distinct branches demonstrates that the self-interacting 3-form field provides a unified framework capable of accommodating different physical configurations—phantom or partially canonical—depending on how the consistency condition is resolved, thereby enlarging the landscape of viable regular black-hole mimickers.

\section{Constraints on the bounce parameter $h_{0}$ from black-hole shadow observations}
\label{sec:shadow}

The Event Horizon Telescope (EHT) has provided unprecedented measurements of the shadow cast by the supermassive black hole Sgr A$^{*}$ at the centre of our galaxy~\cite{EventHorizonTelescope:2022wkp}. These observations offer a powerful tool to constrain deviations from the standard Schwarzschild and Kerr metrics, and therefore to test the viability of alternative compact object solutions such as the black-bounce geometries constructed in this work. The shadow boundary is determined by the photon sphere---the surface of unstable circular null orbits---which depends solely on the spacetime geometry and is therefore a direct probe of the metric functions. In this section, we use the EHT shadow data to place observational bounds on the 3-form parameter $h_{0}$, which controls the size of the bounce and the strength of the Kalb--Ramond field, thereby establishing the range of parameter values for which the solutions are phenomenologically viable.

We restrict our analysis to the asymptotically flat branch of our Universe ($r>0$, with $A(r_{O})\approx 1$ at the observer's position). The relevant observational data for Sgr A$^{*}$, including the mass and distance measurements from the Keck and VLTI surveys, are collected in Table~\ref{tabI}. These measurements provide the essential calibration for translating the observed angular size of the shadow into the dimensionless quantity $r_{\rm sh}/M$ that we compute theoretically.

\begin{table}[t]
	\centering
	\caption{Mass and distance measurements for Sgr A$^{*}$.}
	\label{tabI}
	\begin{tabular}{c c c c}
		\hline\hline
		Survey & $M\,(\times 10^{6}M_{\odot})$ & $r_{O}\;({\rm kpc})$ & Reference \\
		\hline
		Keck 
		& $3.951 \pm 0.047$ 
		& $7.953 \pm 0.050 \pm 0.032$ 
		& \cite{Do:2019txf} \\
		
		VLTI 
		& $4.297 \pm 0.012 \pm 0.040$ 
		& $8.277 \pm 0.009 \pm 0.033$ 
		& \cite{GRAVITY:2020gka} \\
		\hline\hline
	\end{tabular}
\end{table}

\subsection{Shadow radius of a static, spherically symmetric black bounce}

For a static, spherically symmetric metric of the form~\eqref{metric}, the photon sphere---the surface of unstable circular null orbits that defines the boundary of the black-hole shadow---is determined by the condition
\begin{equation}
	\frac{dA(r)}{dr}\,\Sigma^{2}(r) - A(r)\,\frac{d\Sigma^{2}(r)}{dr} = 0.
	\label{ps}
\end{equation}
This equation expresses the requirement that a null geodesic at constant $r$ be circular, which is only possible when the radial effective potential has an extremum. Solving Eq.~\eqref{ps} yields the photon-sphere radius $r_{\rm ps}$. The shadow radius, as measured by a distant observer, is then obtained by tracing the photon trajectory from the photon sphere to the observer's location. For an observer at infinity, this gives the simple relation
\begin{equation}
	r_{\rm sh} = \sqrt{\frac{\Sigma^{2}(r_{\rm ps})}{A(r_{\rm ps})}}.
	\label{sh}
\end{equation}
This formula has been widely employed in the literature to constrain alternative compact objects using EHT data, and it provides a direct link between the spacetime geometry and the observable shadow size.

The EHT collaboration has reported the following observational constraints on the shadow radius of Sgr A$^{*}$~\cite{EventHorizonTelescope:2022wkp}:
\begin{align}
	1\sigma\; \text{confidence:}&\quad 4.55 \lesssim \frac{r_{\rm sh}}{M} \lesssim 5.22,\\
	2\sigma\; \text{confidence:}&\quad 4.21 \lesssim \frac{r_{\rm sh}}{M} \lesssim 5.56.
\end{align}
These intervals represent the current precision of shadow measurements and serve as the benchmark against which we test our solutions.

\subsection{Constraints on the first solution (Path A)}

For the first solution obtained in Section~\ref{sec:epsilon} (Path A), with the phantom scalar field $\epsilon(r)=-1$ and the standard Simpson--Visser areal radius $\Sigma(r)=\sqrt{r^{2}+h_{0}^{2}}$, the photon-sphere equation~\eqref{ps} admits a remarkably simple analytic solution. Substituting the metric function~\eqref{a1} and the profile $\Sigma(r)$ into Eq.~\eqref{ps}, the equation simplifies dramatically, and the photon-sphere radius is found to coincide exactly with the Schwarzschild value,
\begin{equation}
	\frac{r_{\rm ps}}{M} = 3,
\end{equation}
independently of the bounce parameter $h_{0}$. This result is noteworthy: despite the presence of the 3-form field and the associated modifications to the metric function $A(r)$, the location of the photon sphere remains unaffected by $h_{0}$. This can be traced to the fact that, for this particular combination of $\Sigma(r)$ and $A(r)$, the additional terms in the photon-sphere condition cancel identically.

Substituting $r_{\rm ps}=3M$ into Eq.~\eqref{sh} and using the metric function~\eqref{a1} with $\Lambda=0$ (asymptotic flatness), we obtain the shadow radius in closed form:
\begin{equation}
	\frac{r_{\rm sh}}{M} = \sqrt{\frac{2h_{1}^{3}}{h_{1}\!\left(2 - \dfrac{3\pi}{h_{1}}\right) + 6\arctan\!\left(\dfrac{3}{h_{1}}\right)}},
\end{equation}
where we have introduced the dimensionless parameter $h_{1} \equiv h_{0}/M$, which measures the bounce scale in units of the black-hole mass.

The shadow radius as a function of $h_{0}$ is plotted in the left panel of Fig.~\ref{shfig}. The curve exhibits a monotonic dependence on the bounce parameter, allowing a straightforward translation of the EHT bounds into constraints on $h_{0}$. Requiring the solution to be consistent with the $2\sigma$ EHT bounds on the Sgr A$^{*}$ shadow restricts the bounce parameter to the interval
\begin{equation}
	0 \le \frac{h_{0}}{M} \le 1.49.
\end{equation}
Within this range, the solution possesses an event horizon, and the BB describes a regular black hole whose shadow is compatible with current observations. The Schwarzschild limit $h_{0}\to 0$ is included in this interval, confirming that the standard GR prediction lies within the $2\sigma$ contour. For $h_{0}>1.49$, the shadow radius falls outside the $2\sigma$ confidence interval, and the solution would be in tension with the EHT measurements.

\subsection{Constraints on the second solution (Path B)}

\begin{figure*}[htbp!]
	\centering
	\includegraphics[width=0.495\textwidth]{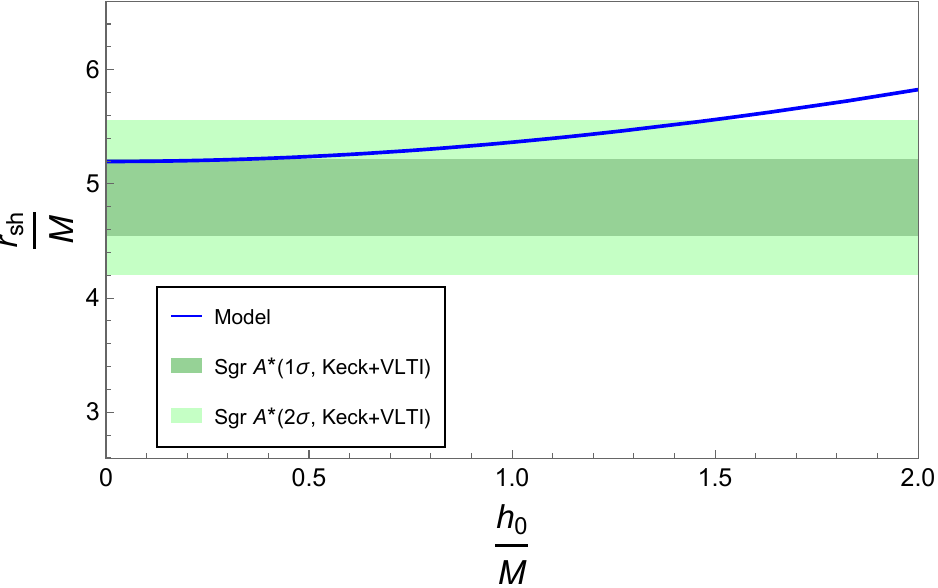}
	\includegraphics[width=0.495\textwidth]{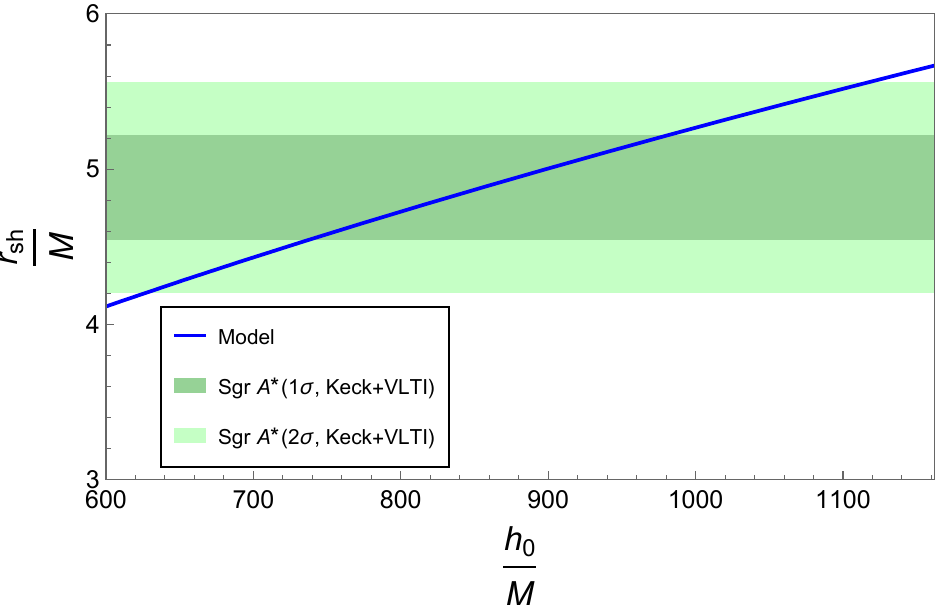}
	\caption{Shadow radius $r_{\rm sh}/M$ as a function of the dimensionless bounce parameter $h_{0}/M$. \textbf{Left panel:} First solution (Path A, $\epsilon=-1$, $\Sigma(r)=\sqrt{r^{2}+h_{0}^{2}}$). The horizontal shaded bands indicate the $1\sigma$ (dark) and $2\sigma$ (light) EHT constraints for Sgr A$^{*}$. The solution is compatible with observations for $h_{0}/M \in [0,\,1.49]$. \textbf{Right panel:} Second solution (Path B, deformed $\Sigma(r)$ with $k_{1}=0.9$). Compatibility with the $2\sigma$ bound requires $h_{0}/M \in [629.56,\,1117.29]$.}
	\label{shfig}
\end{figure*}

For the second solution obtained in Section~\ref{sec:potential} (Path B), with the deformed areal radius $\Sigma(r)$ given by Eq.~\eqref{Sigma2} and the metric function~\eqref{a2}, the photon-sphere equation~\eqref{ps} does not admit an analytic solution due to the increased algebraic complexity introduced by the deformation parameter $k_{1}$. We therefore solve Eq.~\eqref{ps} numerically, scanning the parameter space to determine $r_{\rm ps}$ for each value of $h_{0}$ and subsequently computing the shadow radius via Eq.~\eqref{sh}. The result of this numerical analysis is displayed in the right panel of Fig.~\ref{shfig}.

Imposing consistency with the $2\sigma$ EHT bounds on the Sgr A$^{*}$ shadow yields the constraint
\begin{equation}
	629.56 \le \frac{h_{0}}{M} \le 1117.29.
\end{equation}
The BB possesses an event horizon throughout this interval. The allowed range for $h_{0}$ is significantly larger than for the first solution---by approximately three orders of magnitude---reflecting the profound impact of the areal radius deformation on the exterior geometry. In the second solution, the parameter $k_{1}$ alters the relationship between the bounce scale and the shadow size, effectively shifting the observationally viable region to much larger values of $h_{0}$. This demonstrates that the two solution paths lead to observationally distinguishable predictions for the bounce scale, and that future improvements in shadow precision could potentially discriminate between them.

\subsection{Discussion}

The shadow constraints derived above demonstrate that both families of BB solutions sourced by the self-interacting 3-form field are compatible with current EHT observations of Sgr A$^{*}$, provided the bounce parameter $h_{0}$ lies within the respective allowed intervals. The first solution admits a compact range of $h_{0}$ that includes the Schwarzschild limit, indicating that the 3-form corrections to the shadow are relatively mild for sub-Planckian bounce scales. The second solution, by contrast, requires a much larger bounce scale, reflecting the fact that the deformed areal radius significantly alters the relationship between the bounce parameter and the observable shadow size. This difference can be traced to the distinct roles played by the kinetic coupling function $\epsilon(r)$ and the areal radius deformation parameter $k_{1}$ in shaping the exterior geometry: in Path A, the phantom nature of the scalar field constrains the bounce to be small, whereas in Path B, the partially canonical scalar field allows for a wider range of viable configurations.

Future improvements in the precision of shadow measurements, driven by next-generation instruments and extended observational campaigns by the EHT, may further tighten these constraints and potentially distinguish between the two solution paths. Moreover, extending this analysis to rotating BB solutions---which are more realistic astrophysically---would provide additional observables, such as the shadow asymmetry and the photon ring structure, that could offer even more stringent tests of the 3-form framework.

\section{Conclusion and outlook}
\label{sec:conclusion}

In this work, we have investigated a novel class of black-bounce (BB) solutions sourced by a self-interacting 3-form field minimally coupled to general relativity and a scalar field. The 3-form field, which naturally arises in string theory, supergravity, and various cosmological contexts, provides a rich and versatile matter sector capable of supporting regular geometries that interpolate smoothly between black holes and traversable wormholes. By exploiting the Hodge duality between a 3-form and a 1-form in four dimensions, we reduced the field equations to a tractable system of ordinary differential equations.

A central methodological advance of this work is that the black-bounce solutions are obtained through the direct integration of the equations of motion, rather than through the reconstruction approach commonly employed in the BB literature. In the standard framework, the metric functions---most notably the Simpson--Visser areal radius $\Sigma(r)=\sqrt{r^{2}+h_{0}^{2}}$---are postulated \emph{a priori}, and the Einstein equations are then solved algebraically to determine the matter content required to support the prescribed geometry. While computationally efficient, this reconstruction procedure does not guarantee that the resulting matter fields descend from a well-defined fundamental action, nor does it establish that the geometry arises as a natural solution of the coupled field equations. In contrast, the methodology developed here begins with a complete action principle comprising the 3-form field, the scalar field, and their self-interactions, and proceeds to integrate the coupled gravitational and matter field equations directly. The black-bounce geometry emerges organically from the dynamics of the theory, without any \emph{ad hoc} geometric ansatz. To the best of our knowledge, this represents the first example of a BB solution in general relativity derived in this fully self-consistent manner, establishing a new standard of rigor in the construction of regular bouncing geometries and demonstrating that such spacetimes can arise naturally from well-motivated fundamental fields.

Two complementary solution paths emerged from the consistency condition that ensures the 3-form Lagrangian is a genuine function of the invariant $H_{2}=H_{abc}H^{abc}$. In Path A, we fixed the kinetic coupling function to $\epsilon(r)=-1$, corresponding to a globally phantom scalar field of the type commonly employed in BB constructions. This choice led to a metric function with a distinctive arctangent dependence, an areal radius given by the standard Simpson--Visser profile $\Sigma(r)=\sqrt{r^{2}+h_{0}^{2}}$, and a 3-form Lagrangian that is linear in $H_{2}$. The solution reduces to Schwarzschild--(anti-)de Sitter spacetime in the limit $h_{0}\to 0$, confirming that the 3-form parameter $h_{0}$ controls the deviation from general relativity and plays the role of the bounce scale.

In Path B, we adopted a different strategy: we left the kinetic coupling function $\epsilon(r)$ free and instead solved the consistency equation for the scalar potential $V(r)$. This approach yielded a constant 3-form Lagrangian and, remarkably, a partially canonical scalar field---one that is phantom only in the interior region, near the bounce, and becomes canonical outside the event horizon. To achieve this behavior, we introduced a deformed areal radius $\Sigma(r) = \sqrt{r^{2}+h_{0}^{2}} - k_{1}h_{0}^{2}/\sqrt{r^{2}+h_{0}^{2}}$, parametrized by a new constant $k_{1}\in[0,1)$. The metric function in this case involves an inverse hyperbolic cotangent term, reflecting the distinct algebraic structure induced by the deformation. The emergence of a partially canonical scalar field within pure general relativity is a particularly significant result: in the standard BB framework sourced by non-linear electrodynamics, canonical scalar fields supporting regular solutions have only been obtained in modified theories of gravity, such as $f(R)$ gravity~\cite{deSousaSilva:2026kvp,Silva:2025fqj}. Here, the self-interacting 3-form field provides the additional degrees of freedom necessary to achieve this configuration while remaining within the realm of Einstein's theory.

Both families of solutions are globally regular, as confirmed by the finiteness of the Kretschmann scalar throughout the entire spacetime, including at the bounce $r=0$ and in the asymptotic limits $r\to\pm\infty$. They exhibit an asymmetric horizon structure, with the positive and negative radial branches possessing distinct causal properties---a feature inherited from the asymmetric distribution of the 3-form energy density. The solutions are asymptotically de Sitter, anti-de Sitter, or flat, depending on the value of the cosmological constant $\Lambda$, thereby encompassing a wide range of physically relevant configurations.

To assess the phenomenological viability of our solutions, we constrained the bounce parameter $h_{0}$ using the Event Horizon Telescope observations of the shadow of Sgr A$^{*}$. For the first solution (Path A), compatibility with the $2\sigma$ EHT bounds requires $0 \le h_{0}/M \le 1.49$, a compact interval that includes the Schwarzschild limit. For the second solution (Path B, with $k_{1}=0.9$), the allowed range is significantly larger, $629.56 \le h_{0}/M \le 1117.29$, reflecting the profound impact of the areal radius deformation on the shadow size. The fact that both families admit observationally viable parameter ranges demonstrates that the 3-form BB framework is not only mathematically consistent but also compatible with current astrophysical data.

The results presented in this work open several promising avenues for future investigation. First, the analysis can be extended to rotating BB solutions. Incorporating angular momentum is essential for realistic astrophysical modeling, and the interplay between the 3-form field, the scalar sector, and rotation may yield new classes of regular solutions with richer horizon structures and more complex shadow morphologies. The shadow of a rotating 3-form BB would exhibit an asymmetry that could be tested against future EHT observations, providing stronger constraints on the model parameters.

Second, the thermodynamic properties of these solutions deserve detailed study. The modified horizon structure induced by the 3-form field will affect the Hawking temperature, the entropy, and the evaporation rate. In particular, the presence of a bounce may lead to stable remnants, offering a possible resolution of the information paradox within the 3-form framework. Investigating the thermodynamic stability and phase structure of these solutions would further elucidate their physical interpretation.

Third, the dynamics of test particles and accretion disks around 3-form BBs should be explored. The modified effective potential will shift the innermost stable circular orbit, alter the epicyclic frequencies, and affect the quasi-periodic oscillations observed in X-ray binaries. Comparing these predictions with data from missions such as \textit{NICER} and \textit{XRISM} could provide independent constraints on the 3-form parameter $h_{0}$ and the deformation constant $k_{1}$.

Fourth, the 3-form framework can be applied to other regular geometries beyond the Simpson--Visser family. The direct integration method employed here is general and can accommodate different field configurations, potentially leading to new classes of regular black holes, wormholes, and horizonless compact objects. Exploring this broader landscape may reveal connections with other areas of gravitational physics, such as the hoop conjecture, the weak gravity conjecture, and the swampland program.

Finally, the cosmological implications of the 3-form BB solutions warrant attention. The same 3-form field that sources the bounce on astrophysical scales may also play a role in the early Universe, contributing to inflationary dynamics or dark energy. Embedding these solutions in a cosmological context could provide a unified description of regular compact objects and the large-scale evolution of the Universe.

In summary, this work establishes the self-interacting 3-form field as a powerful and flexible matter source for constructing regular black-bounce solutions in general relativity. The fact that these solutions are obtained through direct integration of the equations of motion, rather than by postulating the geometry and reconstructing the matter content, represents a significant methodological advance in the BB program. The two solution paths explored here---one leading to a phantom scalar field, the other to a partially canonical one---illustrate the richness of the 3-form framework and its ability to accommodate diverse physical configurations. The observational consistency of both families with current EHT shadow data, combined with the novel feature of a partially canonical scalar field emerging within Einstein's theory, positions the 3-form BB as a compelling candidate for extending the standard black-hole paradigm. We hope that the results presented here will stimulate further research at the intersection of modified gravity, string-inspired field theory, and strong-field astrophysics, and that the 3-form black bounce will serve as a valuable laboratory for testing fundamental physics in the era of precision gravitational-wave and black-hole imaging experiments.

\acknowledgments{FSNL acknowledges support from the Funda\c{c}\~{a}o para a Ci\^{e}ncia e a Tecnologia (FCT) Scientific Employment Stimulus contract with reference CEECINST/00032/2018, and funding through the research grant UID/04434/2025.
MER thanks Conselho Nacional de Desenvolvimento Cient\'{\i}fico e Tecnol\'ogico - CNPq, Brazil, for partial financial support.}

\end{document}